\documentclass[aps,prb,twocolumn,showpacs]{revtex4-1}
\pdfoutput=1
\usepackage{graphicx,epsf}
\usepackage{amsmath}
\usepackage{amssymb}
\usepackage{color}
\usepackage{esint}
\usepackage{wasysym}

\newcommand{\bS}{{\bf S}}
\newcommand{\bR}{{\bf R}}
\newcommand{\bq}{{\bf q}}

\newcommand{\bp}{{\bf p}}
\newcommand{\bk}{{\bf k}}

\newcommand{\bn}{{\bf n}}
\newcommand{\taub}{\mbox{\boldmath $\tau $}}

\begin{document}

\title{
 Plaquette valence--bond solid in the square lattice $J_1$--$J_2$
 antiferromagnet Heisenberg model: a bond operator approach
}

\author{R. L. Doretto}
\affiliation{Instituto de F\'isica Gleb Wataghin,
             Universidade Estadual de Campinas,
             13083-859 Campinas, SP, Brazil}

\date{\today}

\begin{abstract}
We study the plaquette valence--bond solid phase of the spin--$1/2$ 
$J_1$--$J_2$ antiferromagnet Heisenberg model on the square lattice
within the bond--operator theory.  
We start by considering four $S = 1/2$ spins on a single plaquette and
determine the bond operator representation for the spin operators in terms of
singlet, triplet, and quintet boson operators. 
The formalism is then applied to the $J_1$--$J_2$ model and an
effective interacting boson model in terms of singlets and triplets is derived. 
The effective model is analyzed within the harmonic approximation and
the previous results of Zhitomirsky and Ueda [Phys. Rev. B {\bf 54},
9007 (1996)] are recovered. 
By perturbatively including cubic (triplet--triplet--triplet and
singlet--triplet--triplet) and quartic interactions, we find that the plaquette
valence--bond solid phase is stable within the parameter region $0.34
< J_2/J_1 < 0.59$, which is narrower than the harmonic one. Differently from the
harmonic approximation, the excitation gap vanishes at both critical
couplings $J_2 = 0.34\, J_1$ and $J_2 = 0.59\, J_1$. 
Interestingly, for $J_2 < 0.48\,J_1$, the excitation gap corresponds to a
singlet--triplet excitation at the $\Gamma$ point while, for $J_2 >
0.48\,J_1$, it is related to a singlet--singlet excitation at the
${\bf X} = (\pi/2,0)$ point of the tetramerized Brillouin zone. 
\end{abstract}

\maketitle

%
%

\section{Introduction}
\label{sec:intro}

\begin{figure*}[t]
\centerline{\includegraphics[width=5.5cm]{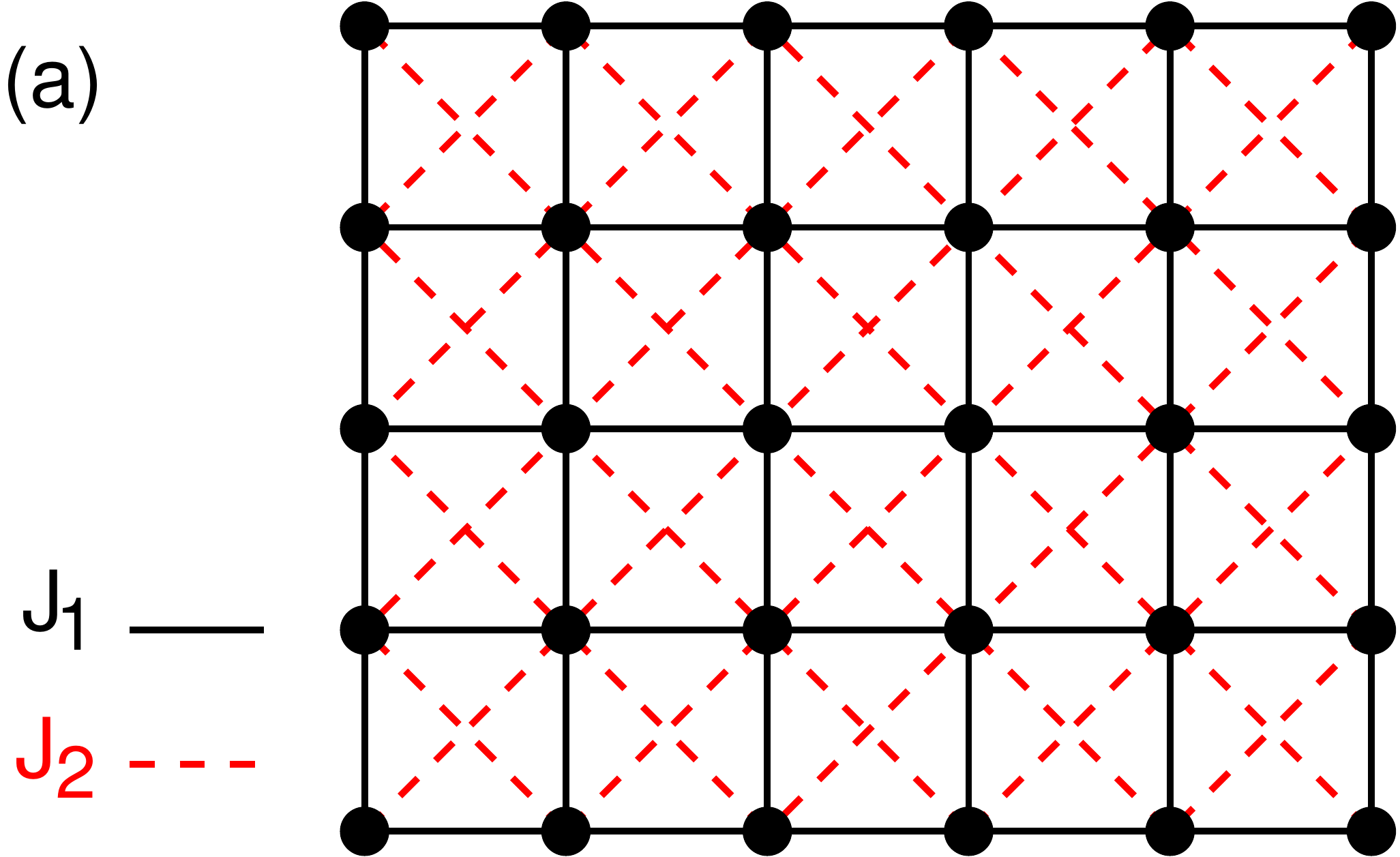}
            \hskip1.5cm
            \includegraphics[width=5.0cm]{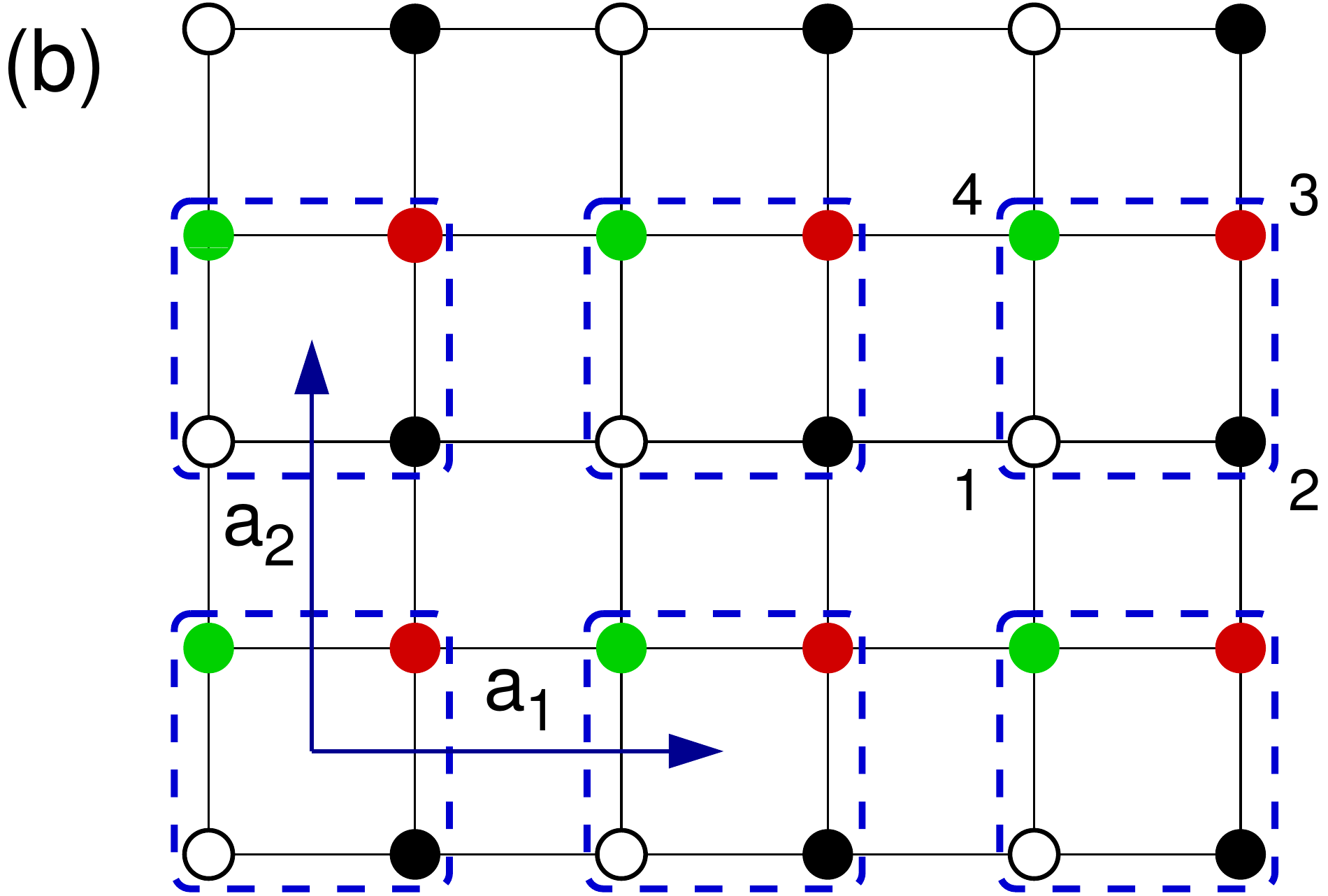}
            \hskip1.5cm
            \includegraphics[width=3.0cm]{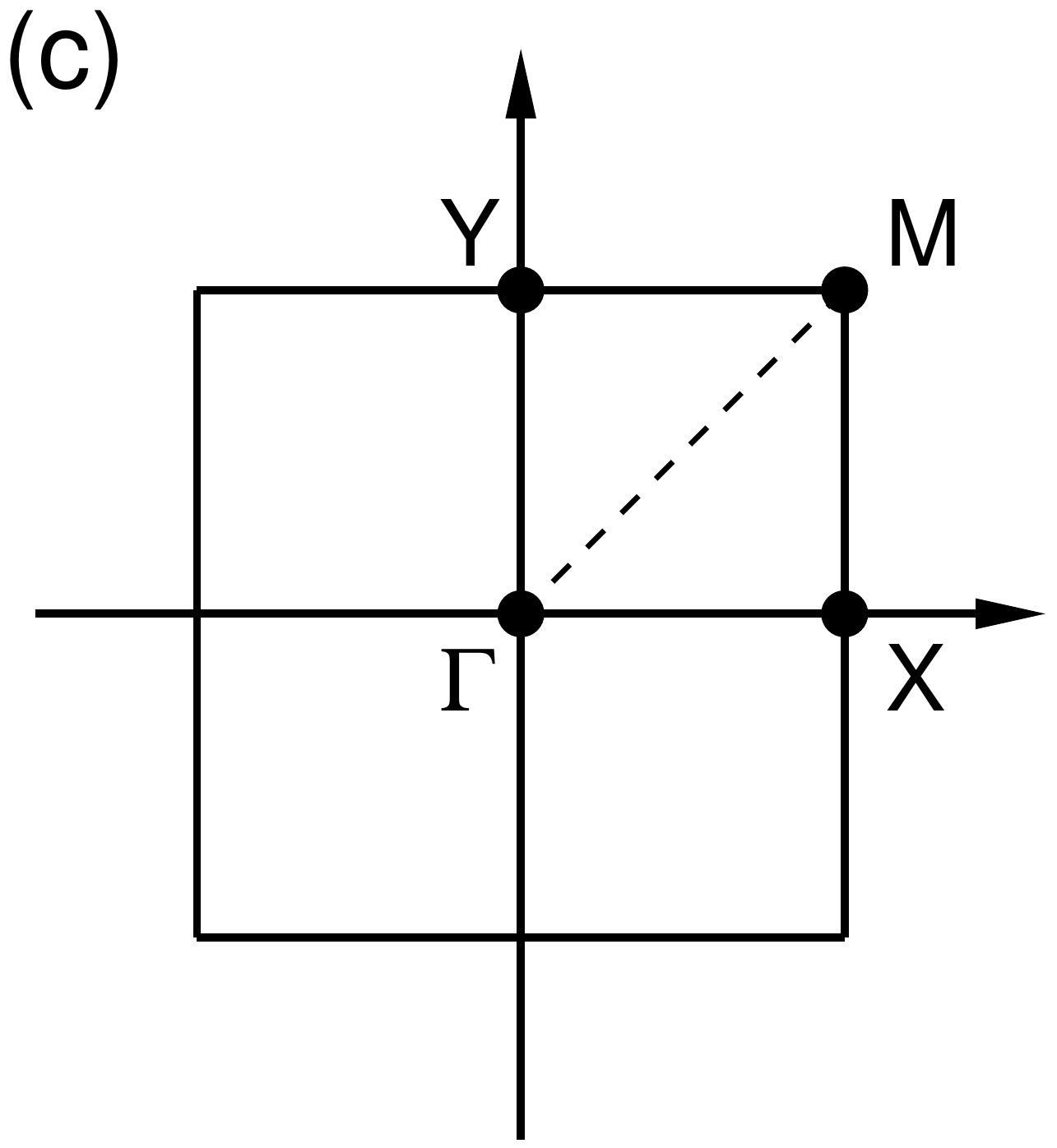}
}
\caption{(Color online) Schematic representation: 
         (a) $J_1$--$J_2$ AFM Heisenberg model \eqref{ham-j1j2}. 
         (b) Plaquette valence bond solid. The dashed blue squares 
         indicate that the 
         spins $\bS^1$ (open circle), $\bS^2$ (black circle), 
         $\bS^3$ (red circle), and $\bS^4$ (green circle)
         form a singlet state. ${\bf a}_1$ and ${\bf a}_2$ are the
         primitive vectors of the tetramerized lattice.
         (c) Brillouin zone of the tetramerized square lattice 
         defined by the plaquettes. 
         Here ${\bf X} = (\pi/2,0)$, ${\bf Y} = (0,\pi/2)$, and 
         ${\bf M} = (\pi/2,\pi/2)$ (the lattice spacing of the 
         original square lattice is set to one).
}
\label{fig:model}
\end{figure*}

Two--dimensional frustrated quantum antiferromagnets have been
receiving a lot of attention in recent years. Here the interplay
between frustration  (dynamic or geometric) and quantum fluctuations
may destroy magnetic long--range order (LRO) yielding to quantum
paramagnetic  (disordered) phases, such as valence bond solids (VBSs)
with broken lattice symmetries or spin liquids, where lattice
symmetries are 
preserved.\cite{review-sachdev,review-frustrated,review-balents}  
An interesting example of a frustrated quantum magnet is the
spin--$1/2$ $J_1$--$J_2$ antiferromagnet (AFM) Heisenberg 
model on the square lattice:\cite{review-j1j2} 
\begin{equation}
 \mathcal{H} = J_1\sum_{\langle ij \rangle} \bS_i\cdot\bS_j 
             + J_2\sum_{\langle\langle ij \rangle\rangle} \bS_i\cdot\bS_j. 
\label{ham-j1j2}
\end{equation}
Here $\bS_i$ is an spin--$1/2$ operator at site $i$ and $J_1 > 0$ and
$J_2 > 0$ are, respectively, the nearest--neighbor and
next--nearest--neighbor exchange couplings as illustrated in
Fig.~\ref{fig:model}(a).  

Several different theoretical approaches have been employed to
study the $J_1$--$J_2$ model in the last few years.
\cite{chandra88,igarashi93,dagotto89,gelfand89,figueirido90,sachdev90,oliveira91,chubukov91,
 read91,schulz92,zhito96,oitamaa96,singh99,kotov99,capriotti00,capriotti01,
 takano03,mambrini06,isaev09,sirker06,darradi08,ralko09,richter10,reuther11,li12,
 mezzacapo12,yu12,gotze12,jiang12,wang13,hu13}
It is now well established that the model has semiclassical N\'eel 
magnetic LRO with ordering wave vector $\bq = (\pi,\pi)$ for 
$J_2 \lesssim 0.4\, J_1$, collinear magnetic LRO with $\bq = (\pi,0)$ 
or $(0,\pi)$ for $J_2 \gtrsim 0.6\, J_1$, and a quantum paramagnetic 
(disordered) phase within the intermediate parameter region 
$0.4 \lesssim J_2/J_1 \lesssim 0.6$. However, the nature of such a 
disordered phase and the quantum phase transition at small $J_2$ 
are still under debate.
These two issues are mainly associated with the fact that  
large--scale quantum Monte Carlo simulations can not 
be used here due to the so-called sign problem.\cite{sign-problem}

\begin{figure}[b]
\centerline{
            \includegraphics[width=4.0cm]{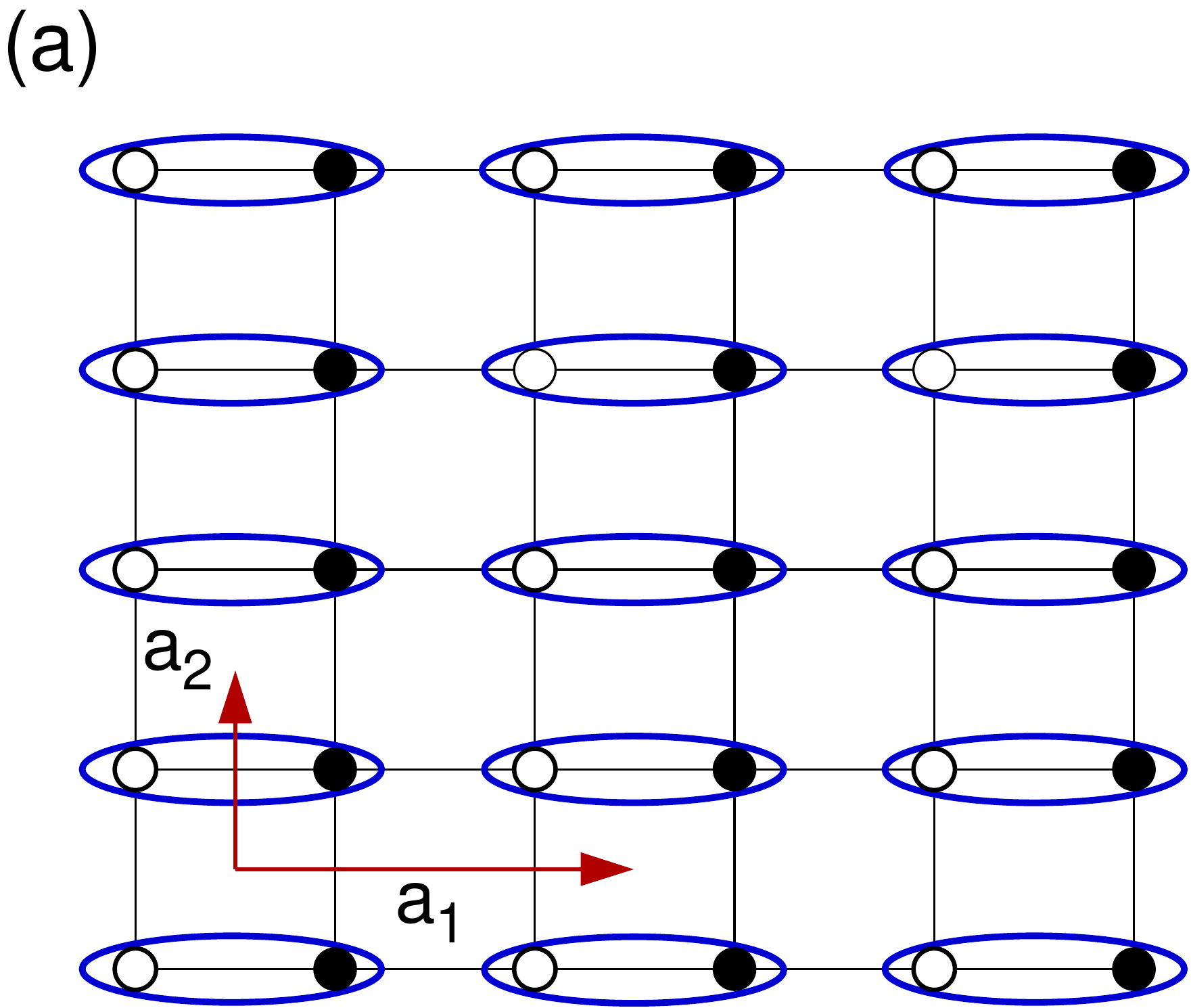}
            \hskip0.5cm
            \includegraphics[width=4.0cm]{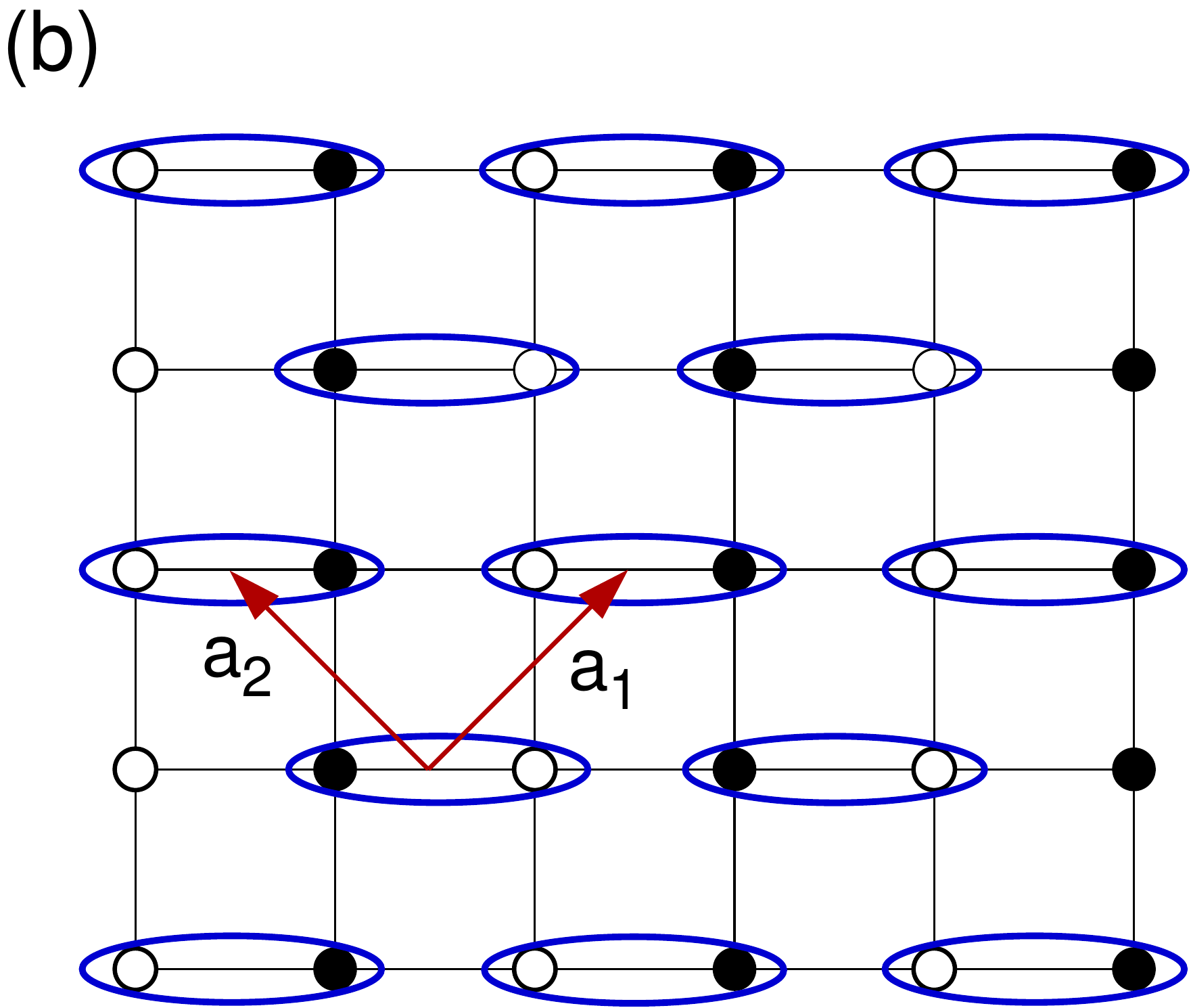}
}
\caption{(Color online) Schematic representation:  
         (a) Columnar and (b) staggered valence bond solids. 
         The blue ellipses indicate 
         that the spins $\bS^1$ (open circle) and $\bS^2$ (filled circle)
         form a singlet state.
         ${\bf a}_1$ and ${\bf a}_2$ are the
         primitive vectors of the dimerized lattices.
}
\label{fig:dimer}
\end{figure}

Different proposals have been made for the ground state of the 
disordered phase of the $J_1$--$J_2$ model:  
a columnar VBS [Fig.~\ref{fig:dimer}(a)], where both translational and 
  rotational lattice symmetries are broken,\cite{sachdev90,kotov99,singh99}
a plaquette VBS [Fig.~\ref{fig:model}(b)], where only the
 translational lattice symmetry is broken,\cite{mambrini06,isaev09,zhito96,
 takano03,yu12,capriotti00}
a mixed columnar--plaquette VBS,\cite{ralko09}
and gapless spin--liquids.\cite{capriotti01,wang13,hu13}
More recently, evidences for a gapped ${\rm Z}_2$ 
spin--liquid\cite{li12,mezzacapo12,jiang12} have also been found.

About the quantum phase transitions: 
while there are strong indications\cite{darradi08,gotze12,yu12,li12} 
that a first--order quantum phase transition takes place at $J_2 \approx 0.6\,J_1$
(the boundary between the quantum paramagnetic and the collinear phases),
it is still not clear whether a first--order\cite{sirker06} 
or a continuous\cite{darradi08,isaev09,gotze12,yu12,li12} quantum 
phase transition occurs at $J_2 \approx 0.4\,J_1$ 
(the boundary between the N\'eel and the quantum paramagnetic phases).  
If a VBS phase sets in within the magnetic disorder region, the
former scenario is in agreement with the Landau--Ginzburg framework
(the N\'eel and the VBS phases are characterized by two different 
order parameters)
while the latter is in favor of the so-called deconfined quantum
criticality.\cite{senthil04}
A candidate theory for a possible continuous quantum phase transition
between a  ${\rm Z}_2$ spin--liquid and a N\'eel phase is recently 
proposed in Ref.~\onlinecite{moon12}.

We should also mention that, more recently, the $J_1$--$J_2$ model 
on the {\sl honeycomb} lattice has also been studied. 
Here the main motivation are quantum Monte Carlo results\cite{meng10} 
for the half--filled honeycomb Hubbard model which provide some 
evidences for a gapped spin--liquid phase within intermediate values
of the on--site repulsion $U$. 
Density matrix renormalization group (DMRG) calculations have been
performed on the honeycomb lattice $J_1$--$J_2$ model and it is found that as
$J_2/J_1$ increases, a N\'eel phase, a plaquette and a dimerized 
VBS phases set in.\cite{ganesh13,zhu13,gong13}
Similar results are reported in Ref.~\onlinecite{bishop13}, where the
coupled cluster method is employed.

An useful approach to describe VBS phases of a Heisenberg model is 
the bond--operator theory introduced by Sachdev and Bhatt.\cite{sachdev90} 
Such a formalism can be seen as the analog of the Holstein--Primakoff 
representation, but here we consider fluctuations above a quantum 
paramagnetic ground state instead of a (semiclassical) state with 
magnetic LRO. 
The formalism developed in Ref.~\onlinecite{sachdev90} is
appropriate to describe dimerized phases, such as the columnar 
[Fig.~\ref{fig:dimer}(a)] and staggered [Fig.~\ref{fig:dimer}(b)]
VBSs.\cite{sachdev90,kotov99}   
A generalized method suitable for describing tetramerized phases,
such as the plaquette VBS [Fig.~\ref{fig:model}(b)], was later
introduced by Zhitomirsky and Ueda.\cite{zhito96} 
However, here only a partial bond--operator representation for
the spin operators [in terms of the lowest--energy singlet and 
the triplet (boson) operators] was considered: the high--energy 
singlet and the quintet operators (see below) were neglected.

In this paper, we revisit the work of Zhitomirsky and Ueda\cite{zhito96}
and study the plaquette VBS phase of the $J_1$--$J_2$ model
within the bond--operator theory. 
We derive the full bond--operator representation 
(in terms of singlet, triplet, and quintet boson operators) for
spin--$1/2$ operators on a single plaquette and apply such 
a generalized formalism to the $J_1$--$J_2$ model \eqref{ham-j1j2}. 
Our study is not only restricted to the analysis at the harmonic (mean--field) 
level of an effective boson model in terms of the lowest--energy 
singlet and the triplet operators as done in Ref.~\onlinecite{zhito96}, 
but we also include the high--energy singlet operator and go beyond the
harmonic approximation: cubic (singlet--triplet--triplet 
and triplet--triplet--triplet) and quartic interactions are perturbatively
considered.
Our main motivations are a series of
results\cite{mambrini06,isaev09,zhito96,takano03,yu12,capriotti00}
which indicates the stability of the plaquette VBS phase and 
a recent study\cite{doretto12} concerning a dimerized phase of a
triangular lattice Heisenberg AFM, where we show that cubic 
(triplet--triplet--triplet)
interactions have an important role in the determination of the 
excitation spectrum of such a frustrated quantum magnet.

\begin{figure}[b]
\centerline{\includegraphics[width=6.5cm]{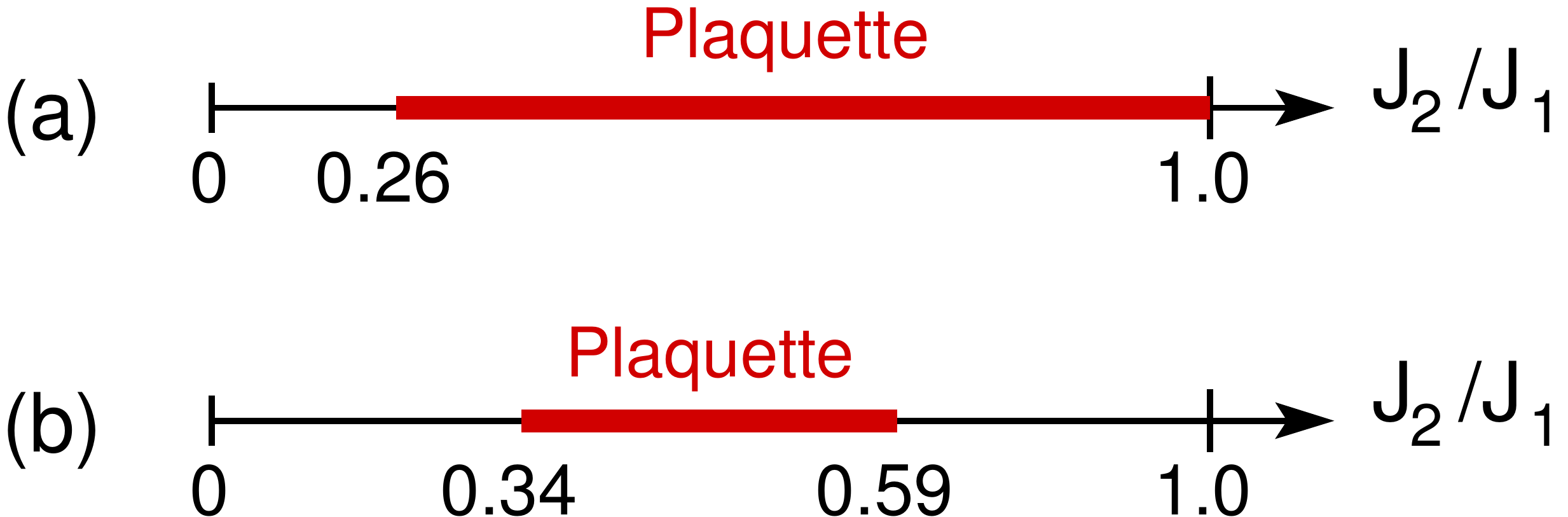}
}
\caption{(Color online) Region of stability of the plaquette VBS phase
  of the $J_1$--$J_2$ model \eqref{ham-j1j2}
  as obtained from bond--operator theory within 
  (a) the harmonic and (b) the cubic--quartic approximations. 
}
\label{fig:phase-diag}
\end{figure}

\subsection{Overview of the results}

We calculate the ground state energy [Fig.~\ref{fig:egs}(b)] and the
dispersion relation of the singlet and triplet excitations
(Fig.~\ref{fig:disp}) of the plaquette VBS phase within the
(mean--field) harmonic and the cubic--quartic approximations. In the latter, 
cubic and quartic interactions are perturbatively added to the harmonic results.
Our main findings are the following: \\
(a) Harmonic approximation. The plaquette phase is stable within the
parameter region $0.26 < J_2/J_1 < 1.00$, see
Fig.~\ref{fig:phase-diag}(a). The excitation gap [Fig.~\ref{fig:gap}(a)] is
always finite and it is related to a singlet--triplet excitation 
(triplet gap) for $J_2 < 0.82\,J_1$ and 
a singlet--singlet one (singlet gap) for $J_2 > 0.82\,J_1$. \\
(b) Cubic--quartic approximation. The region of stability of the plaquette
phase is $0.34 < J_2/J_1 < 0.59$ [Fig.~\ref{fig:phase-diag}(b)] with the excitation
gap vanishing at both critical couplings $J_2 = 0.34\,J_1$ and
$J_2 = 0.59\,J_1$ [Fig.~\ref{fig:gap}(b)].  
For $J_2 > 0.48\,J_1$, the excitation gap is no
longer associated with a singlet--triplet excitation at the $\Gamma$
point, but with a singlet--singlet one at the ${\bf X} = (\pi/2,0)$
point of the tetramerized Brilluoin zone [see Fig.~\ref{fig:model}(c)].
The decay rates of the singlet and triplet excitations are also
obtained [see Figs.~\ref{fig:decay} and \ref{fig:decay2}].

The reader not interested in the technical details may skip 
Secs. II -- V and go straight to Sec. VI.

\subsection{Outline}

Our paper is organized as follows: In Sec.~\ref{sec:bondop}, we generalize 
the (dimer) bond--operator formalism\cite{sachdev90} for the case of four 
spins $S=1/2$ on a single plaquette. In Sec.~\ref{sec:boson-model}, we apply the 
generalized bond--operator representation to the $J_1$--$J_2$ model  
and derive an effective model in terms of singlet and triplet boson operators.
Sec.~\ref{sec:harmonic} is devoted to the analysis of the effective boson model
in the harmonic approximation. The ground state energy and the dispersion relations 
of the singlet and triplet excitations are calculated. In Sec.~\ref{sec:cubic}, we 
consider cubic (singlet--triplet--triplet and triplet--triplet--triplet) interactions 
in second--order perturbation theory and quartic ones in the (no self--consistent) 
Hartree--Fock approximation and calculate the corrections to the harmonic results 
(cubic--quartic approximation). We compare our results with previous ones and discuss their
implications for the $J_1$--$J_2$ model in Sec.~\ref{sec:discussion}. 
Our findings are summarized in the last section. Some details
of the calculations discussed in the main part can be found in the five Appendixes.

\section{Bond operator representation}
\label{sec:bondop}

In Ref.~\onlinecite{sachdev90}, a bond--operator representation
for two spins $S = 1/2$ in a dimer is introduced. In this section, we
consider the case of four spins $S = 1/2$ in a plaquette and
develop a bond--operator representation for the spin operators in terms
of singlet, triplet, and quintet (boson) operators.  
We should mention that such a formalism was already 
discussed in Refs.~\onlinecite{zhito96} and \onlinecite{ueda07}
but, in that case, the high--energy singlet state $|s_1\rangle$ and the quintet 
states $|d_0\rangle$, $|d_2\rangle$, and $|d_\alpha\rangle$ (see below) 
were not considered. 
As far as we know, this is the first time that the complete bond--operator 
representation for spins in a plaquette is derived.  

\begin{figure}[t]
\centerline{\includegraphics[width=4.5cm]{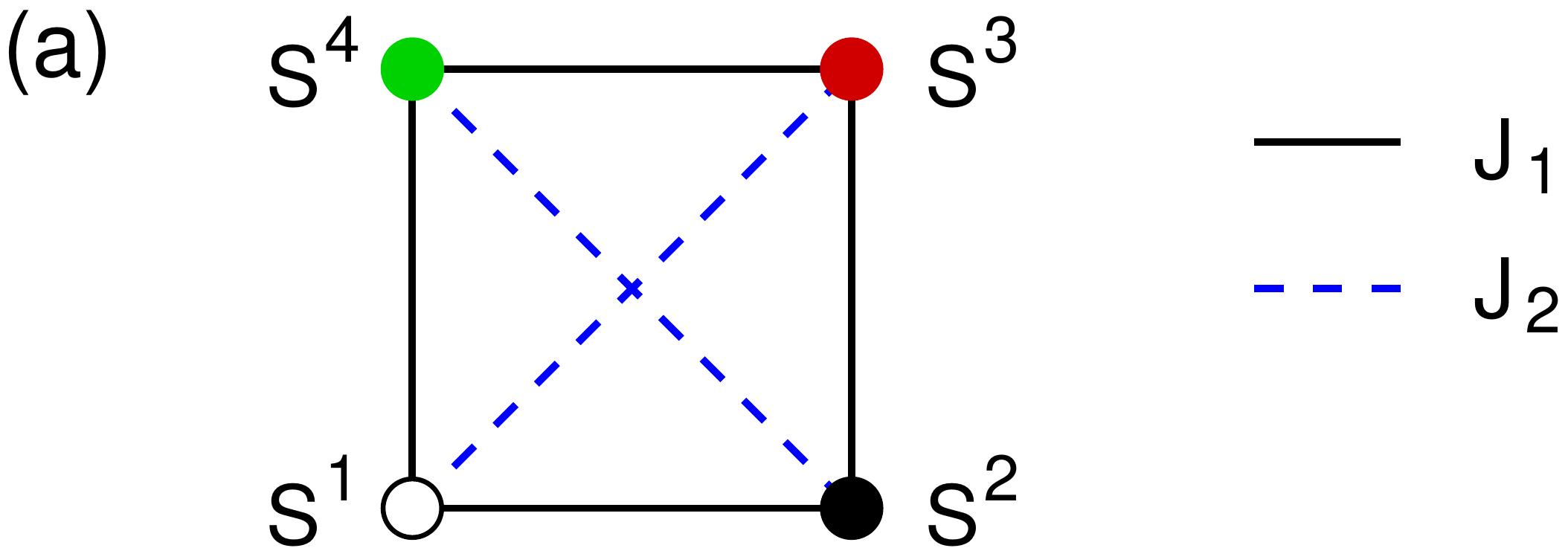}}
            \vskip0.7cm
\centerline{\includegraphics[width=7.1cm]{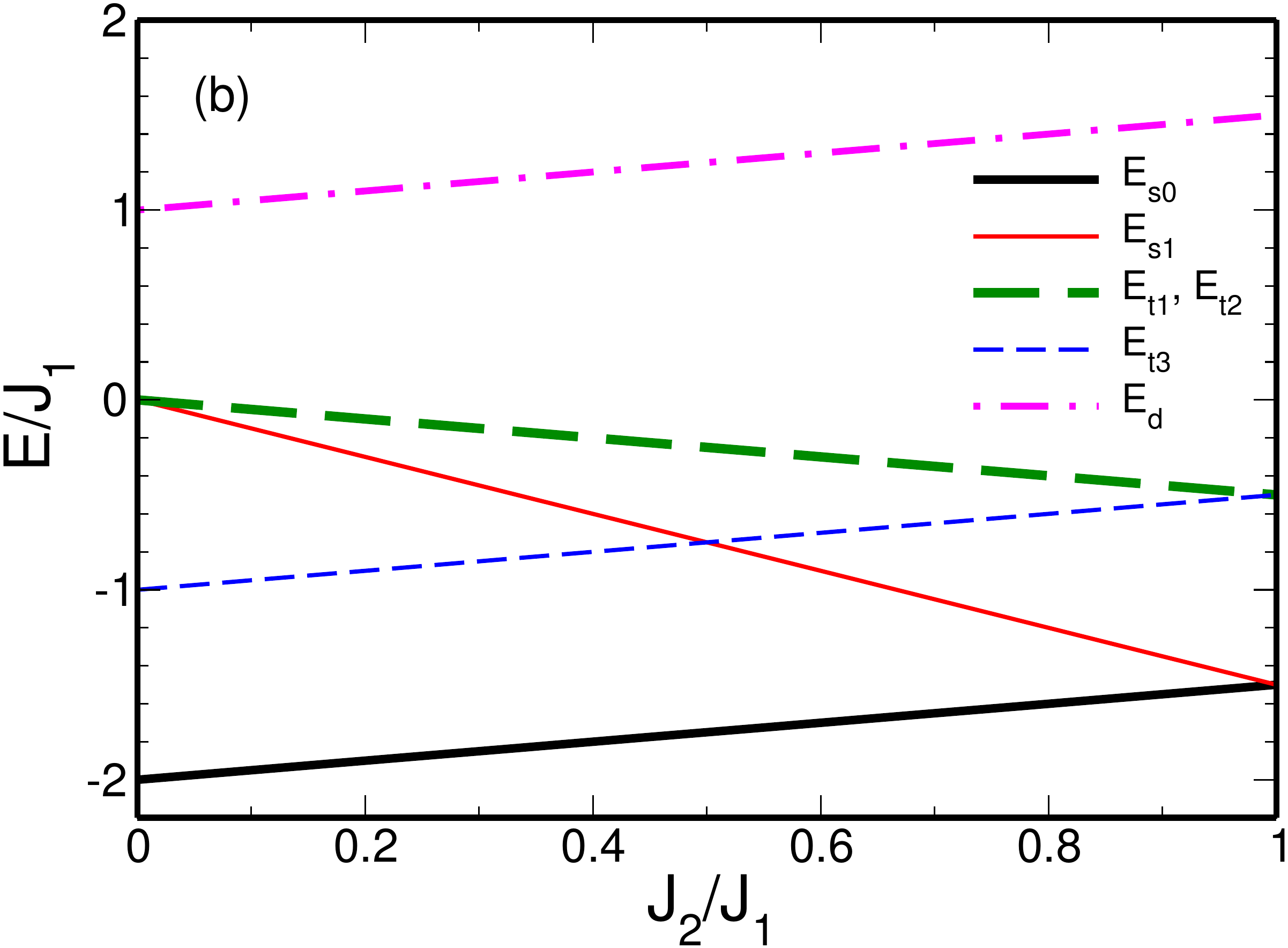}
}
\caption{(Color online)(a) Schematic representation of the spin--$1/2$
         $J_1$--$J_2$ AFM Heisenberg model \eqref{ham-j1j2} on a
         single plaquette, Eq.~\eqref{ham-j1j2-plaq}.
         (b) Eigenvalues \eqref{eigenvalues} of the Hamiltonian 
         \eqref{ham-j1j2-plaq} as a function of $J_2/J_1$: $E_{s0,s1}$,
         $E_{t1,t2,t3}$, and $E_d$ are, respectively, the energies of 
         the singlet, triplet and quintet states.}
\label{fig:plaquette}
\end{figure}

\subsection{Single plaquette}

Let us consider the Heisenberg model \eqref{ham-j1j2} restricted to four spins
in a single plaquette as illustrated in Fig.~\ref{fig:plaquette}(a):
\begin{equation}
 \mathcal{H}_{plaq} = J_1 \left( \bS^1 + \bS^3 \right)\cdot\left( \bS^2 + \bS^4 \right) 
             + J_2 \left( \bS^1\cdot\bS^3 + \bS^2\cdot\bS^4 \right).
\label{ham-j1j2-plaq}
\end{equation}
It is easy to show that the eigenvalues of the Hamiltonian \eqref{ham-j1j2-plaq} 
are given by
\begin{eqnarray}
  E_{s0} &=& -2J_1 + \frac{1}{2}J_2, \;\;\;\;\;\;\;
  E_{s1}  = -\frac{3}{2}J_2,
\label{eigenvalues} \\
&& \nonumber \\
  E_{t1,t2} &=& -\frac{1}{2}J_2,    \;\;\;\;
  E_{t3}    = -J_1 + \frac{1}{2}J_2,\;\;\;\; 
  E_d      = J_1 + \frac{1}{2}J_2. 
\nonumber
\end{eqnarray}
The behaviour of the spectrum  as a function of $J_2/J_1$ is shown in 
Fig.~\ref{fig:plaquette}(b). For $J_2 < J_1$, the ground state is given 
by the singlet state $|s_0\rangle$ whose energy is $E_{s0}$. 
There are four excited energy levels:
$E_{s1}$ is the eigenvalue related to the singlet state $|s_1\rangle$.
$E_{t1} = E_{t2}$ is the energy of the six triplet states $|t_{1,\alpha}\rangle$
and $|t_{2,\alpha}\rangle$ with $\alpha = x,y,z$ while $E_{t3}$ is the energy 
of the three triplet states $|t_{3,\alpha}\rangle$. Finally, $E_d$ is the
eigenvalue associated with the five quintet states $|d_0\rangle$, $|d_2\rangle$, 
and $|d_\alpha\rangle$. 
Note that the excitation gap is associated with a singlet--triplet transition
(triplet gap) for $J_2 < 0.5J_1$ and with a singlet--singlet one
(singlet gap) for $J_2 > 0.5J_1$.
We refer the reader to Appendix \ref{ap:hilbert}
for the explicit expressions of the singlet, triplet and quintet states in 
terms of the 16 states $|\uparrow\,\uparrow\,\uparrow\,\uparrow \rangle$,
$|\downarrow\,\uparrow\,\uparrow\,\uparrow \rangle$, 
$|\uparrow\,\downarrow\,\uparrow\,\uparrow \rangle$, 
$\ldots$ , 
$|\downarrow\,\downarrow\,\downarrow\,\downarrow \rangle$.

\subsection{Boson operators}

As discussed in the previous section, the Hilbert space of four  
spins $S=1/2$ ($\mathbf S^1$, $\mathbf S^2$, $\mathbf S^3$, and $\mathbf S^4$)
in a single plaquette is made out of 16 states: two singlet, nine triplet, 
and five quintet states. We can introduce a set of boson operators which 
creates these states out of a fictitious vacuum $|0\rangle$, namely,
\begin{eqnarray}
 | s_0 \rangle &=& s_0^\dagger |0\rangle, \;\;\;\;\;\;\; 
 | s_1 \rangle  =  s_1^\dagger |0\rangle, \;\;\;\;\;\;\; 
 | t_{a,\alpha} \rangle = t_{a,\alpha}^\dagger |0\rangle,
\nonumber \\
&&  \label{def-bosons} \\
 | d_0 \rangle &=& d_0^\dagger |0\rangle, \;\;\;\;\;\;\;
 | d_2 \rangle = d_2^\dagger |0\rangle,   \;\;\;\;\;\;\; 
 | d_\alpha \rangle = d_\alpha^\dagger |0\rangle,
\nonumber
\end{eqnarray}
with $a=1,2,3$ and $\alpha = x,y,z$. In order to remove unphysical states 
from the enlarged Hilbert space, the constraint
\begin{equation}
 s^\dagger_0 s_0 + s^\dagger_1 s_1 + \sum_{a,\alpha} t^\dagger_{a,\alpha}t_{a,\alpha}
 + d^\dagger_0d_0 + d^\dagger_2 d_2 + \sum_\alpha d^\dagger_\alpha d_\alpha = 1
\label{constraint}
\end{equation}
should be introduced.

Following the ideas of Ref.~\onlinecite{sachdev90}
for the dimer case, we calculate the matrix elements of each component of 
the four spins operators within the basis 
$\{ | s_0 \rangle, | s_1 \rangle, | t_{a,\alpha} \rangle,
    | d_0 \rangle, | d_2 \rangle, | d_\alpha \rangle    \} $, i.e., we
determine $\langle s_0 | S^\mu_\alpha | s_1 \rangle$, 
$\langle s_0 | S^\mu_\alpha | t_{a,\beta} \rangle$, $\ldots$, with $\mu = 1,2,3,4$. 
Based on the obtained results, one concludes that the three components of the four spin 
operators $\mathbf S^\mu$ can be written in terms of boson operators
$s$, $t$, and $d$ as
\begin{widetext}
\begin{eqnarray}
S^\mu_\alpha &=& \frac{1}{2\sqrt{3}} ( \pm t^\dagger_{a,\alpha} -(-1)^\mu \sqrt{2}t^\dagger_{3,\alpha} )s_0 
                + {\rm h.c.}
                \pm \frac{1}{2}t^\dagger_{b,\alpha}s_1 + {\rm h.c.} 
             -(-1)^\mu  \frac{1}{2\sqrt{3}} ( t^\dagger_{3,\alpha} \pm(-1)^\mu\sqrt{2}t^\dagger_{a,\alpha} )(
                             \cos\theta_\alpha d_0 + \sin\theta_\alpha d_2) +{\rm h.c.}
\nonumber \\
&& \nonumber \\
&+&  \frac{1}{2}d^\dagger_\alpha(\sin\theta_\alpha d_0 - \cos\theta_\alpha d_2) + {\rm h.c.}
    - \frac{i}{4}\epsilon^{\alpha\beta\gamma}( 2t^\dagger_{b,\beta}t_{b,\gamma} 
       + t^\dagger_{3,\beta}t_{3,\gamma} - d^\dagger_\beta d_\gamma )    
\nonumber \\
&& \nonumber \\
 &\pm& (-1)^\mu \frac{i}{2\sqrt{2}}\epsilon^{\alpha\beta\gamma}(t^\dagger_{a,\beta}t_{3,\gamma}
        + t^\dagger_{3,\beta}t_{a,\gamma} )    
  \pm  \frac{i}{2\sqrt{2}}I^{\alpha\beta\gamma}(t^\dagger_{a,\beta}d_\gamma - d^\dagger_\beta t_{a,\gamma} )  
       +(-1)^\mu\frac{i}{4}I^{\alpha\beta\gamma}(t^\dagger_{3,\beta}d_\gamma - d^\dagger_\beta t_{3,\gamma} ),  
\label{spin-bondop} 
\end{eqnarray}
\end{widetext}
with $\mu = 1,2,3,4$ and $\alpha,\beta,\gamma = x,y,z$. 
Here,  the upper and lower signs refer respectively to 
$\mu = 1,2$ and $\mu=3,4$,
$(a,b) = (1,2)$ and $(2,1)$ respectively for $\mu = 1,3$ and $\mu = 2,4$,
$\epsilon^{\alpha\beta\gamma}$ is the completely antisymmetric tensor with $\epsilon^{xyz} = 1$, 
$I^{\alpha\beta\gamma} = |\epsilon^{\alpha\beta\gamma}|$ is a symmetric tensor, 
$\theta_x = 2\pi/3$, $\theta_y = 4\pi/3$, and $\theta_z = 0$,
and summation convention over repeated indices is implied.
Similarly, one shows that the Hamiltonian \eqref{ham-j1j2-plaq}
assumes the form
\begin{eqnarray}
  \mathcal{H}_{plaq} &=& E_{s0}s^\dagger_0 s_0 + E_{s1}s^\dagger_1 s_1
                  + E_{t1}\sum_{a=1,2}t^\dagger_{a,\alpha}t_{a,\alpha}
\nonumber \\ 
    &+&    E_{t3}\,t^\dagger_{3,\alpha}t_{3,\alpha}
        + E_d\left( d^\dagger_0d_0 + d^\dagger_2 d_2 +
                    d^\dagger_\alpha d_\alpha \right).
\end{eqnarray}

Since the bond operator representation \eqref{spin-bondop} is quite
involved, it is useful to consider an approximate expansion for
the spin operators $S^\mu_ \alpha$. In particular, neglecting the
high--energy quintet states, Eq.~\eqref{spin-bondop} reduces to
\begin{eqnarray}
S^\mu_\alpha &=& C^\mu_a (t^\dagger_{a,\alpha}s_0 + s^\dagger_0t_{a,\alpha} ) 
  + \bar{C}^\mu_a (t^\dagger_{a,\alpha}s_1 + s^\dagger_1t_{a,\alpha} ) 
\nonumber \\
&& \nonumber \\
    &-&  i\epsilon^{\alpha\beta\gamma} D^\mu_{ab} t^\dagger_{a,\beta}t_{b,\gamma},
\label{spin-bondop3}
\end{eqnarray} 
where $\mu = 1,2,3,4$ and the coefficients $C^\mu_a$, $\bar{C}^\mu_a$, 
and $D^\mu_{ab}$ are given by 
\begin{eqnarray} 
  C^{1/3}_1 &=& C^{2/4}_2 = \pm 1/2\sqrt{3}, \;\;\;\;\;\;\; 
  C^{1/3}_3 = -C^{2/4}_3 = 1/\sqrt{6}, 
\nonumber \\ 
  \bar{C}^{1/3}_2 &=& \bar{C}^{2/4}_1 = \pm 1/2, 
\nonumber \\ 
  D^{2/4}_{11} &=& D^{1/3}_{22} = 1/2, \;\;\;\;\;\;\; 
  D^{2/4}_{33} = D^{1/3}_{33} = 1/4,
\nonumber \\ 
  D^{1/3}_{13} &=& D^{1/3}_{31} = 
  -D^{2/4}_{23} = -D^{2/4}_{32} = \pm1/2\sqrt{2}, 
\label{spin-bondop3-coef}
\end{eqnarray}
and zero otherwise. Eq.~\eqref{spin-bondop3} is quite similar
to the bond operator representation for two spins $S = 1/2$ in a dimer, see e.g.,
Eqs.~(2.2) and (2.3) from Ref.~\onlinecite{sachdev90}.

The bond operator representation \eqref{spin-bondop}  
can be generalized to the lattice case and 
the corresponding Heisenberg model can be expressed in terms of 
the boson operators $s^\dagger_{0,i}$, $s^\dagger_{1,i}$, $t^\dagger_{a,i,\alpha}$, 
$d^\dagger_{0,i}$, $d^\dagger_{2,i}$, and $d^\dagger_{i,\alpha}$.

\begin{figure*}[t]
\centerline{\includegraphics[width=7.8cm]{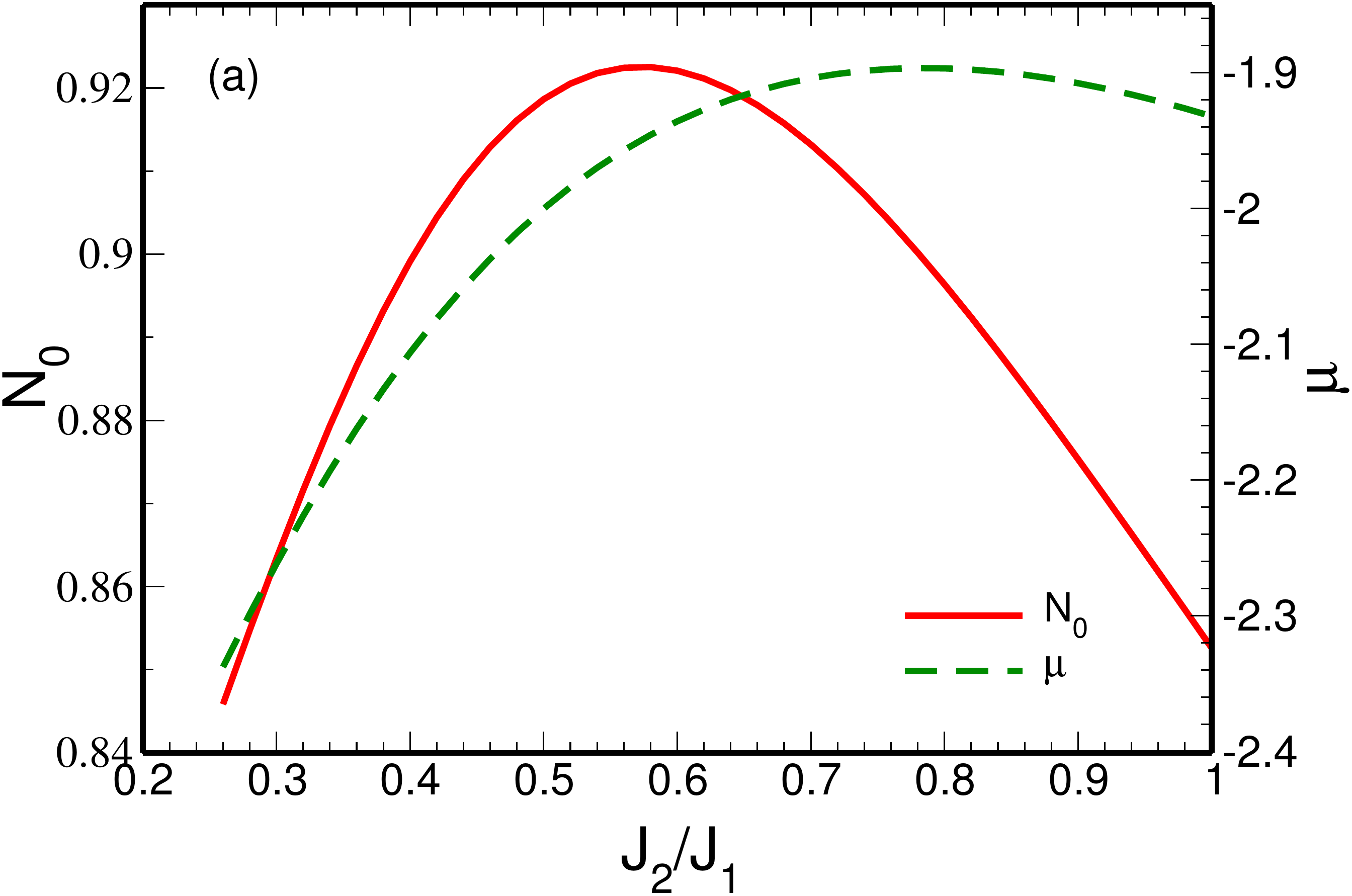}
            \hskip1.7cm
            \includegraphics[width=7.5cm]{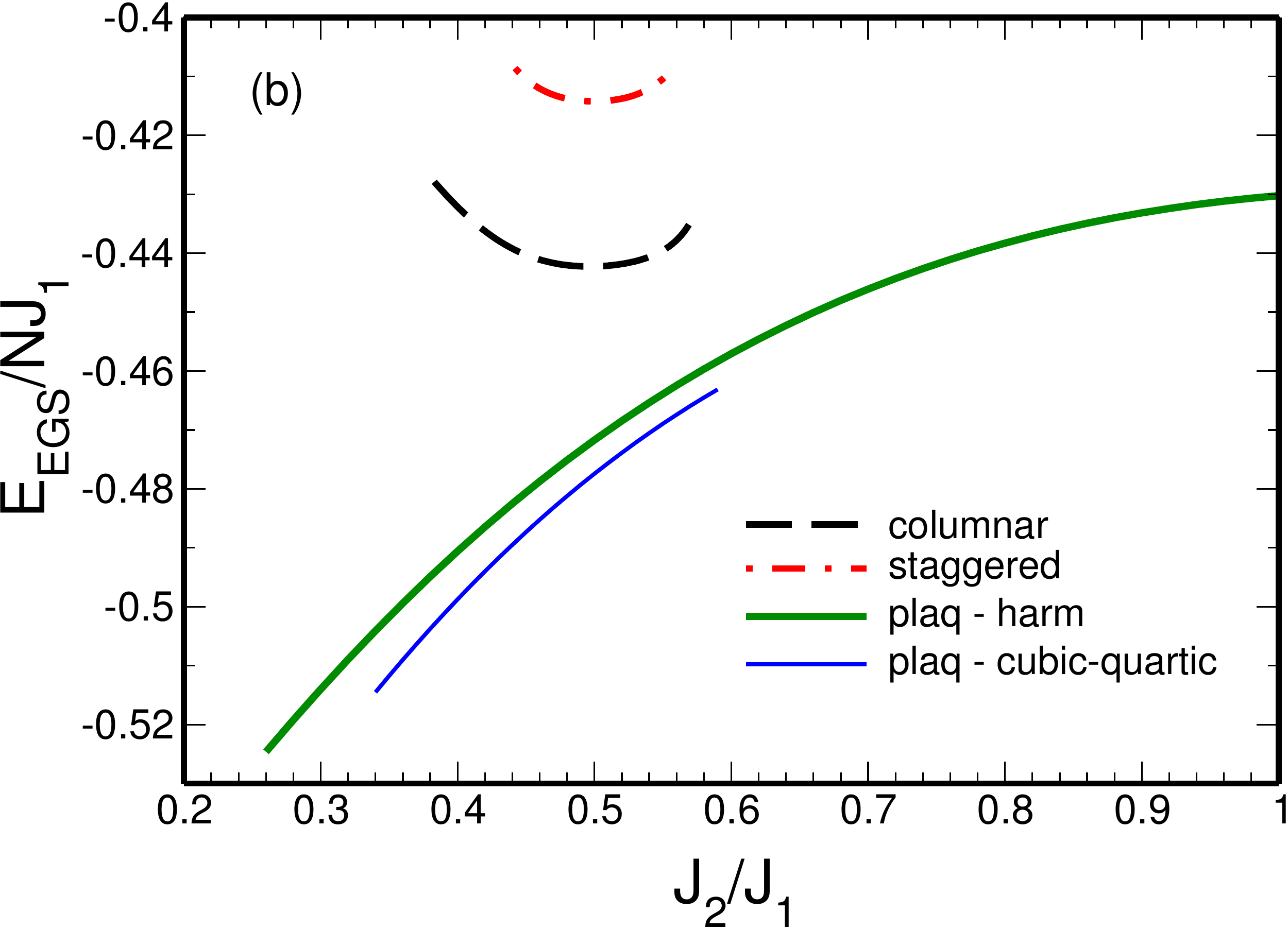}
}
\caption{(Color online) (a) Parameters $N_0$ and $\mu$ [Eqs.~\eqref{self-eq}] as 
          a function of $J_2/J_1$ at the harmonic level. (b) Ground state energies 
          per site as a function of $J_2/J_1$ of the columnar (dashed black line), 
          staggered (dotted-dashed red line), plaquette (thick solid green line) 
          VBSs at the harmonic level [Eq.~\eqref{egs}], and plaquette VBS 
          (thin solid blue line) within
          the cubic--quartic approximation [Eq.~\eqref{total-egs}].}
\label{fig:egs}
\end{figure*}

\section{Effective boson model}
\label{sec:boson-model}

In this section, we apply the bond operator formalism developed above
to study the plaquette VBS phase of the $J_1$--$J_2$ model. The idea
is to map the Heisenberg model \eqref{ham-j1j2} into 
an effective boson model in terms of the singlet $s_{1,i}$ and
the triplet $t_{a,i,\alpha}$ operators.

We start by rewriting the Hamiltonian \eqref{ham-j1j2} in terms of the
underline (tetramerized) square lattice defined by the plaquettes as
shown in Fig.~\ref{fig:model}(b):
\begin{eqnarray}
 \mathcal{H} &=& \sum_i 
       J_1 \left( \bS^1_i + \bS^3_i \right)\cdot\left( \bS^2_i + \bS^4_i \right) 
     + J_2 \left( \bS^1_i\cdot\bS^3_i + \bS^2_i\cdot\bS^4_i \right)
\nonumber \\
&& \nonumber \\
&& + \;J_1\left( \bS^2_i\cdot\bS^1_{i+1} + \bS^3_i\cdot\bS^4_{i+1} 
               + \bS^4_i\cdot\bS^1_{i+2} + \bS^3_i\cdot\bS^2_{i+2} \right)
\nonumber \\
&& \nonumber \\
  && \;+ J_2\left( \bS^2_i\cdot\bS^4_{i+1} + \bS^2_i\cdot\bS^4_{i-2} + \bS^2_i\cdot\bS^4_{i+1-2} 
          \right.
\nonumber \\
&& \left. \right. \nonumber \\
    && \left. \;\;\;\;\; +\;\; 
          \bS^3_i\cdot\bS^1_{i+1} + \bS^3_i\cdot\bS^1_{i+2} + \bS^3_i\cdot\bS^1_{i+1+2}
        \right).
\label{ham-j1j2-plaquette}
\end{eqnarray}
Here, the numbers $1$ and $2$ in the site indices $i+1$, $i+2, \, \ldots$, etc 
respectively indicates the nearest-neighbor vectors
\begin{equation}
 \taub_1 = 2a\hat{x} \;\;\;\;\;\;\; {\rm and} \;\;\;\;\;\;\;
 \taub_2 = 2a\hat{y}
\label{tauvectors}
\end{equation}
with $a$ being the lattice spacing of the {\it original} square
lattice (in the following we set $a=1$). Note that the unit cell
of the underline square lattice has four spins: $\bS^1_i$, $\bS^2_i$, 
$\bS^3_i$ and $\bS^4_i$.
We then substitute Eq.~\eqref{spin-bondop3} generalized to the
lattice case into Eq.~\eqref{ham-j1j2-plaquette}, i.e., we consider
the approximate bond--operator representation where the high--energy
quintets are neglected, and, after some algebra, 
find that the Hamiltonian assumes the general form:
\begin{equation}
  \mathcal{H} = E_0 + \mathcal{H}_{02}
                    + \mathcal{H}_{20} + \mathcal{H}_{30} + \mathcal{H}_{40} 
                    + \mathcal{H}_{21} + \mathcal{H}_{22}.
\label{ham-bond}
\end{equation}
Here $E_0$ is a constant, 
\[
 E_0 = \frac{1}{4}N\left[ N_0E_{s0}  - \mu(N_0 - 1)  \right],
\]
the terms $\mathcal{H}_{nm}$ contain $n$ triplet $t_{a,i,\alpha}$
and $m$ singlet $s_{1,i}$ operators, and the constraint \eqref{constraint} is 
taking into account by adding to the Hamiltonian \eqref{ham-bond} 
the term
\begin{eqnarray}
 && -\mu \sum_i \left(  s^\dagger_{0,i} s_{0,i} + s^\dagger_{1,i} s_{1,i} 
                     + t^\dagger_{a,i,\alpha}t_{a,i,\alpha}  - 1\right)
\nonumber
\end{eqnarray}
with $\mu$ being a Lagrange multiplier.

Within the bond operator formalism, the plaquette VBS state shown in  
Fig.~\ref{fig:model}(b) can be seen as a condensate of the
lowest--energy singlets $s_{0,i}$. In order to implement such a  
(reference) state, we replace
\[
 s^\dagger_{0,i} = s_{0,i} = \langle s^\dagger_{0,i} \rangle =
              \langle s_{0,i} \rangle \rightarrow \sqrt{N_0}
\]
in Eq.~\eqref{ham-bond}. We then end up with an effective 
Hamiltonian solely in terms of the triplet 
$t^\dagger_{a,i,\alpha}$  and the singlet $s^\dagger_{1,i}$ boson operators. 
Both $\mu$ and $N_0$ will be self-consistently determined later. 

\begin{figure*}[t]
\centerline{\includegraphics[width=6.5cm]{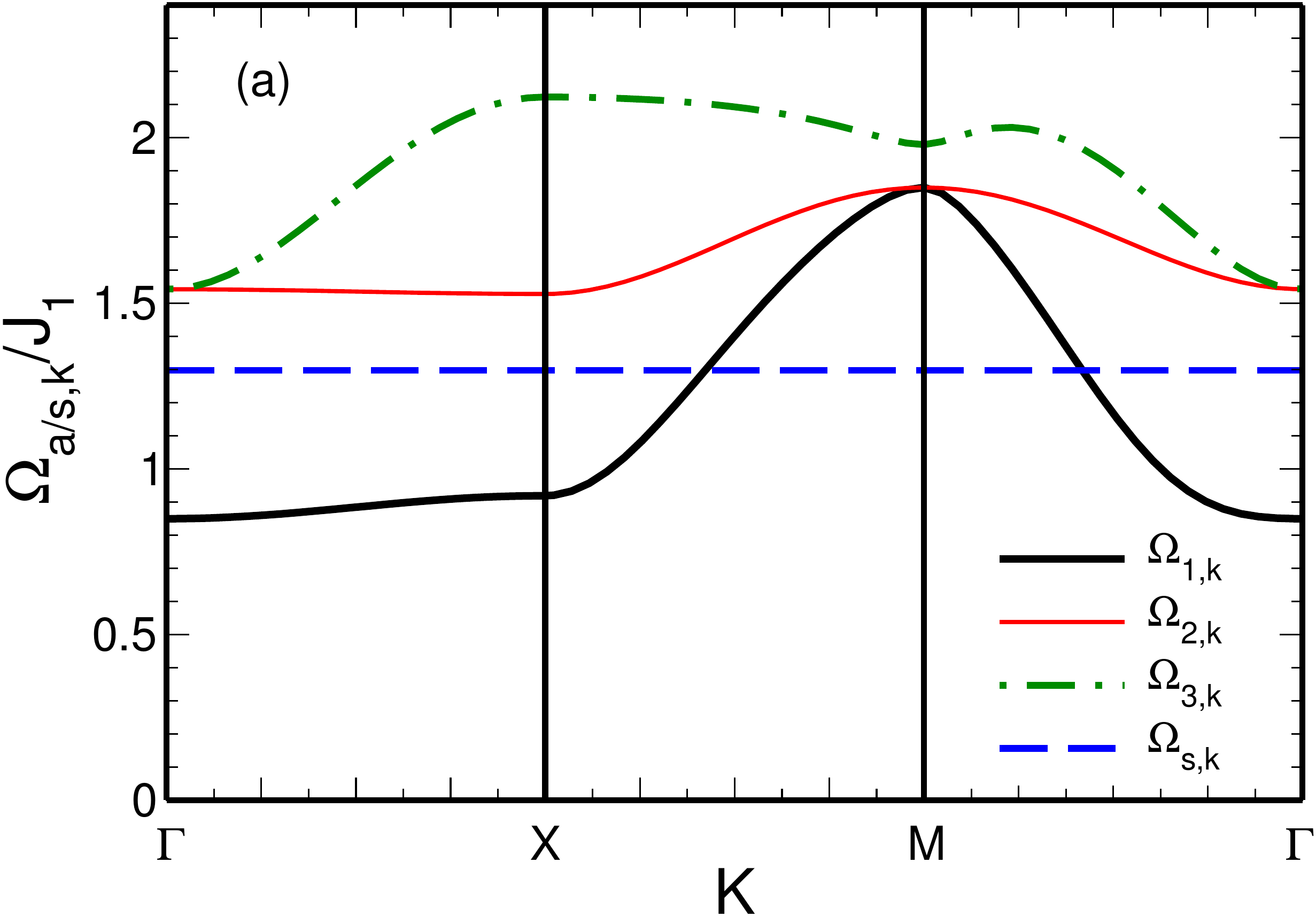}
            \hskip2.0cm
            \includegraphics[width=6.5cm]{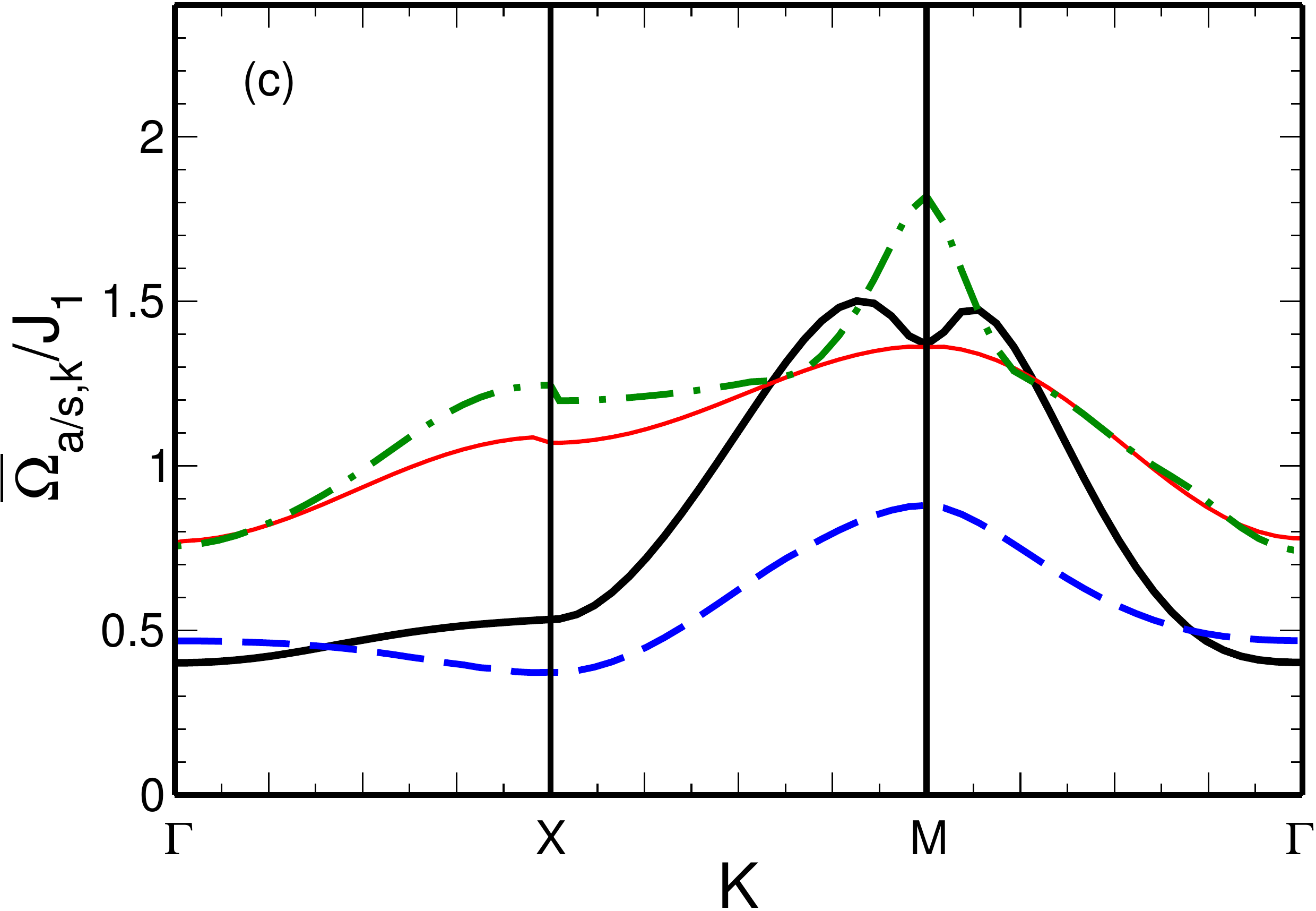}
}
\vskip1.0cm
\centerline{\includegraphics[width=6.5cm]{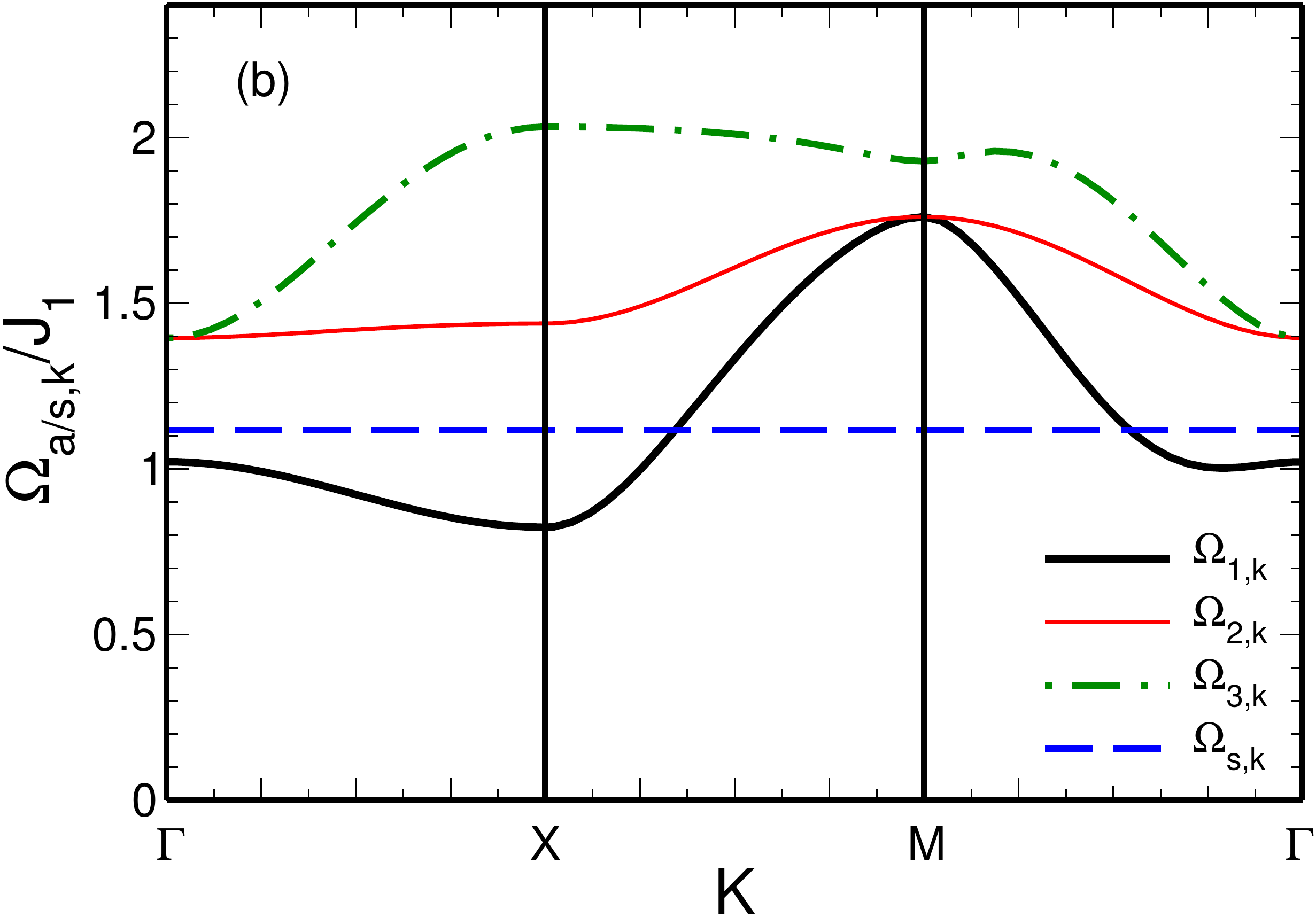}
            \hskip2.0cm
            \includegraphics[width=6.5cm]{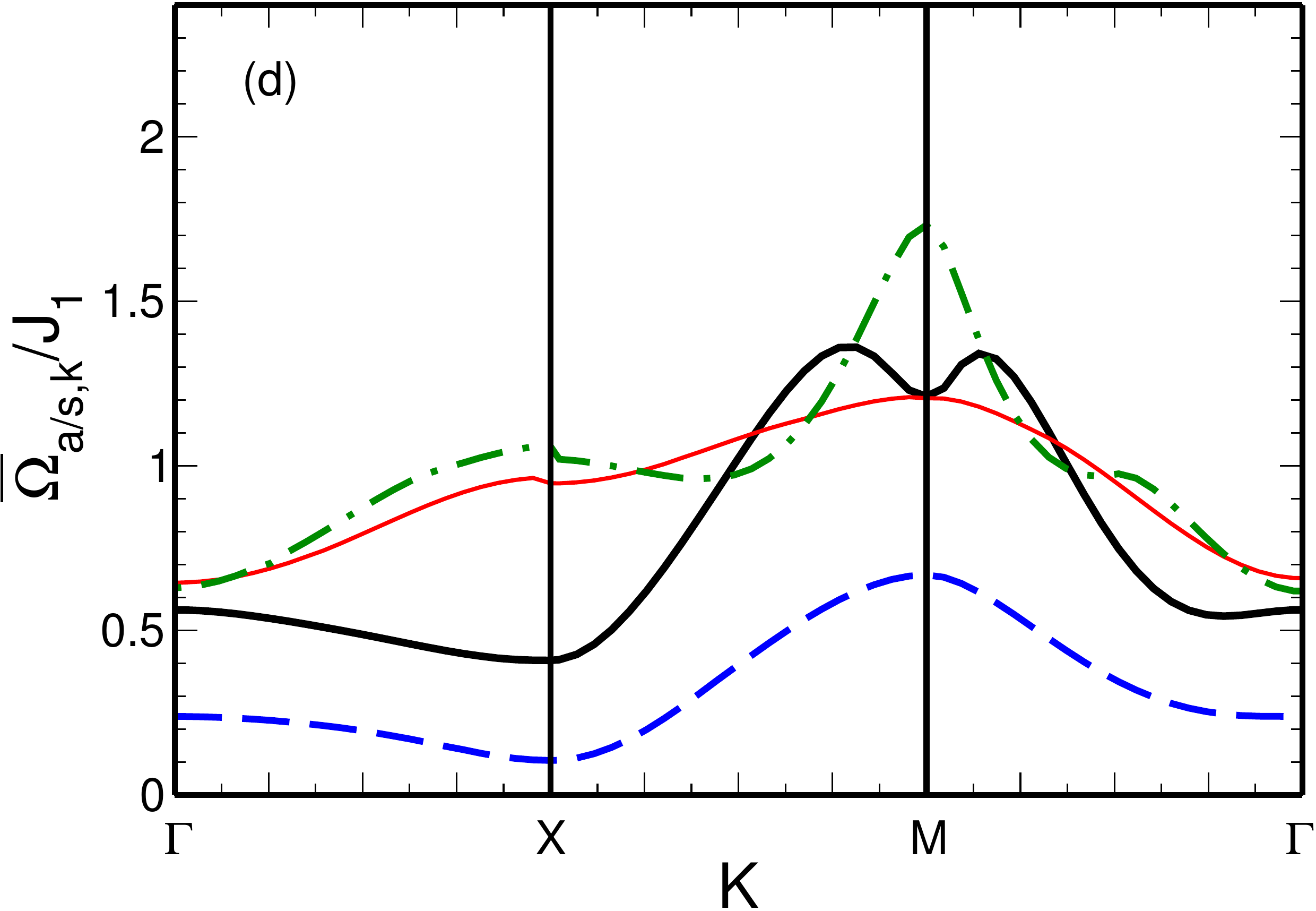}
}
\caption{(Color online) Dispersion relation of the singlet $\Omega_{s,\bk}$ (dashed blue line) 
         and triplet $\Omega_{1,\bk}$ (thick solid black line),
                     $\Omega_{2,\bk}$ (thin solid red line), and
                     $\Omega_{3,\bk}$ (dotted--dashed green line)
          excitations along paths in the tetramerized
          Brillouin zone [Fig.~\ref{fig:model}(c)] at the harmonic level
          for (a) $J_2 = 0.48\,J_1$ and (b) $J_2 = 0.56\,J_1$ and within the 
          cubic--quartic approximation for (c) $J_2 = 0.48\,J_1$ 
          and (d) $J_2 = 0.56\,J_1$.}
\label{fig:disp}
\end{figure*}

Finally, performing a Fourier transform, i.e.,  
\[
  t^\dagger_{a,i,\alpha} = 
     N'^{-1/2}\sum_\bk \exp(-i\bk\cdot\bR_i)t^\dagger_{a,\bk,\alpha},
\]
\[
  s^\dagger_{1,i} = N'^{-1/2}\sum_\bk \exp(-i\bk\cdot\bR_i)s^\dagger_{1,\bk},
\]
where $N' = N/4$ with $N$ being the number of sites of the original
square lattice and
the momentum sums run over the tetramerized Brillouin
zone [Fig.~\ref{fig:model}(c)], we find that in momentum 
space the $\mathcal{H}_{nm}$ terms in Eq.~\eqref{ham-bond} read
\begin{equation}
 \mathcal{H}_{02} = \sum_\bk (E_{s1} - \mu) s^\dagger_{1,\bk}s_{1,\bk}, 
\end{equation}
\begin{equation}
\mathcal{H}_{20} = \sum_\bk A^{ab}_\bk t^\dagger_{a,\bk,\alpha}t_{b,\bk,\alpha}
              + \frac{B^{ab}_\bk}{2} \left( t^\dagger_{a,\bk,\alpha}t^\dagger_{b,\bk,\alpha}
                  + {\rm H.c.}\right),
\label{h20} 
\end{equation}
\begin{equation}
\mathcal{H}_{30} = \frac{\epsilon^{\alpha\beta\gamma}}{\sqrt{N'}}\sum_{\bp,\bk}
                  \xi^{abc}_{\bp-\bk}\;
                  t^\dagger_{a,\bk-\bp,\alpha}t^\dagger_{b,\bp,\beta}t_{c,\bk,\gamma} 
                  + {\rm H.c.},
\label{h30}
\end{equation}
\begin{equation} 
\mathcal{H}_{40} = \frac{\epsilon^{\alpha\beta\gamma}\epsilon^{\alpha\lambda\nu}}{N'}
                  \sum_{\bp,\bq,\bk} \chi^{abcd}_\bk \;
                  t^\dagger_{a,\bp+\bk,\beta}t^\dagger_{b,\bq-\bk,\lambda}t_{c,\bq,\nu}t_{d,\bp\gamma},
\label{h40} 
\end{equation}
\begin{eqnarray}
\mathcal{H}_{21} &=& \frac{1}{\sqrt{N'}}\sum_{\bp,\bk}
                 \left[ \bar{\xi}^{ab}_\bp\;
                  t^\dagger_{a,\bk-\bp,\alpha}t^\dagger_{b,\bp,\alpha}s_{1,\bk}
                  + {\rm H.c.} \right.
\nonumber \\
&& \nonumber \\
 && \left. + \; \bar{\xi}^{ba}_\bp\;
                  s^\dagger_{1,\bk-\bp}t^\dagger_{a,\bp,\alpha}t_{b,\bk,\alpha}
                  + {\rm H.c.} \right],
\label{h21}
\end{eqnarray}
\begin{eqnarray}
\mathcal{H}_{22} &=& \frac{1}{N'}
                  \sum_{\bp,\bq,\bk} \left[ \bar{\chi}^{ab}_\bk \;
                   s^\dagger_{1,\bq+\bk}s_{1,\bp+\bk}t^\dagger_{a,\bp,\alpha}t_{b,\bq\alpha}
                   \right.
\nonumber \\
&& \nonumber \\
 && \left.  + \; \frac{1}{2}\bar{\chi}^{ab}_\bk \;
                   s^\dagger_{1,\bq+\bk}s^\dagger_{1,\bp-\bk}t_{a,\bp,\alpha}t_{b,\bq\alpha}
                   + {\rm H.c.} \right],
\label{h22}
\end{eqnarray}
with $a,b,c = 1,2,3$ and $\alpha,\beta,\gamma = x,y,z$. 
The coefficients $A^{ab}_\bk$, $B^{ab}_\bk$, $\xi^{abc}_{\bk}$,
$\chi^{abcd}_\bk$, $\bar{\xi}^{ba}_\bp$, and $\bar{\chi}^{ab}_\bk$ 
can be found in Appendix~\ref{ap:coef}.

\section{Harmonic approximation}
\label{sec:harmonic}

Let us now study the effective boson model \eqref{ham-bond}
in the lowest--order approximation, the so--called
harmonic approximation. In this case, we neglect $\mathcal{H}_{30}$, 
$\mathcal{H}_{40}$, $\mathcal{H}_{21}$, and $\mathcal{H}_{22}$
and consider
\begin{equation}
  \mathcal{H} \approx E_0 + \mathcal{H}_{02} + \mathcal{H}_{20}.
\label{ham-harmonic}
\end{equation}
Note that the Hamiltonian \eqref{ham-harmonic} is quadratic in the 
singlet $s_{1,\bk}$ and the triplet $t_{a,\bk,\alpha}$ boson operators. 
Moreover, the singlet sector is already diagonalized and decoupled from 
the triplet one. 

\begin{figure*}[t]
\centerline{\includegraphics[width=7.5cm]{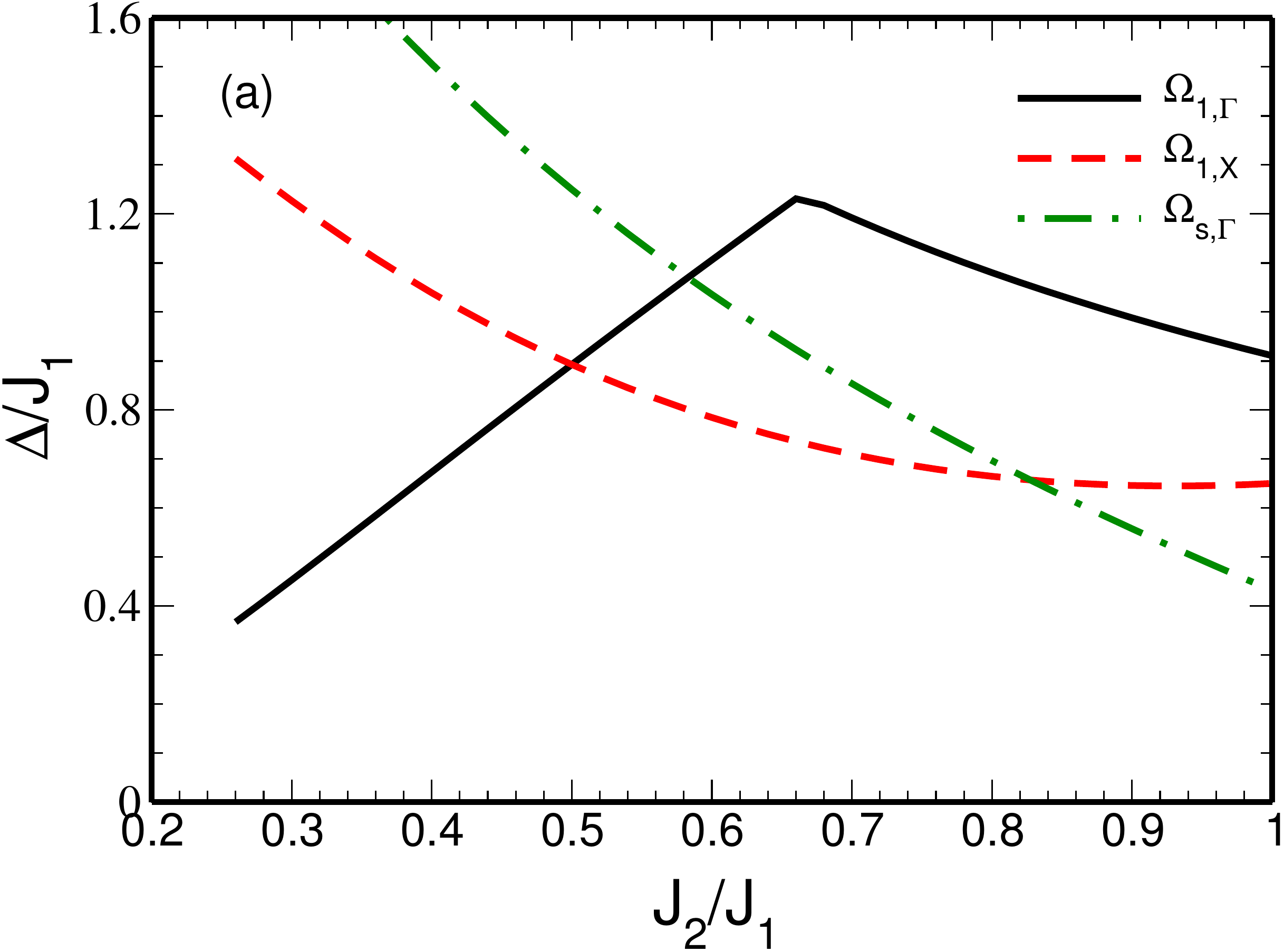}
            \hskip2.0cm
            \includegraphics[width=7.5cm]{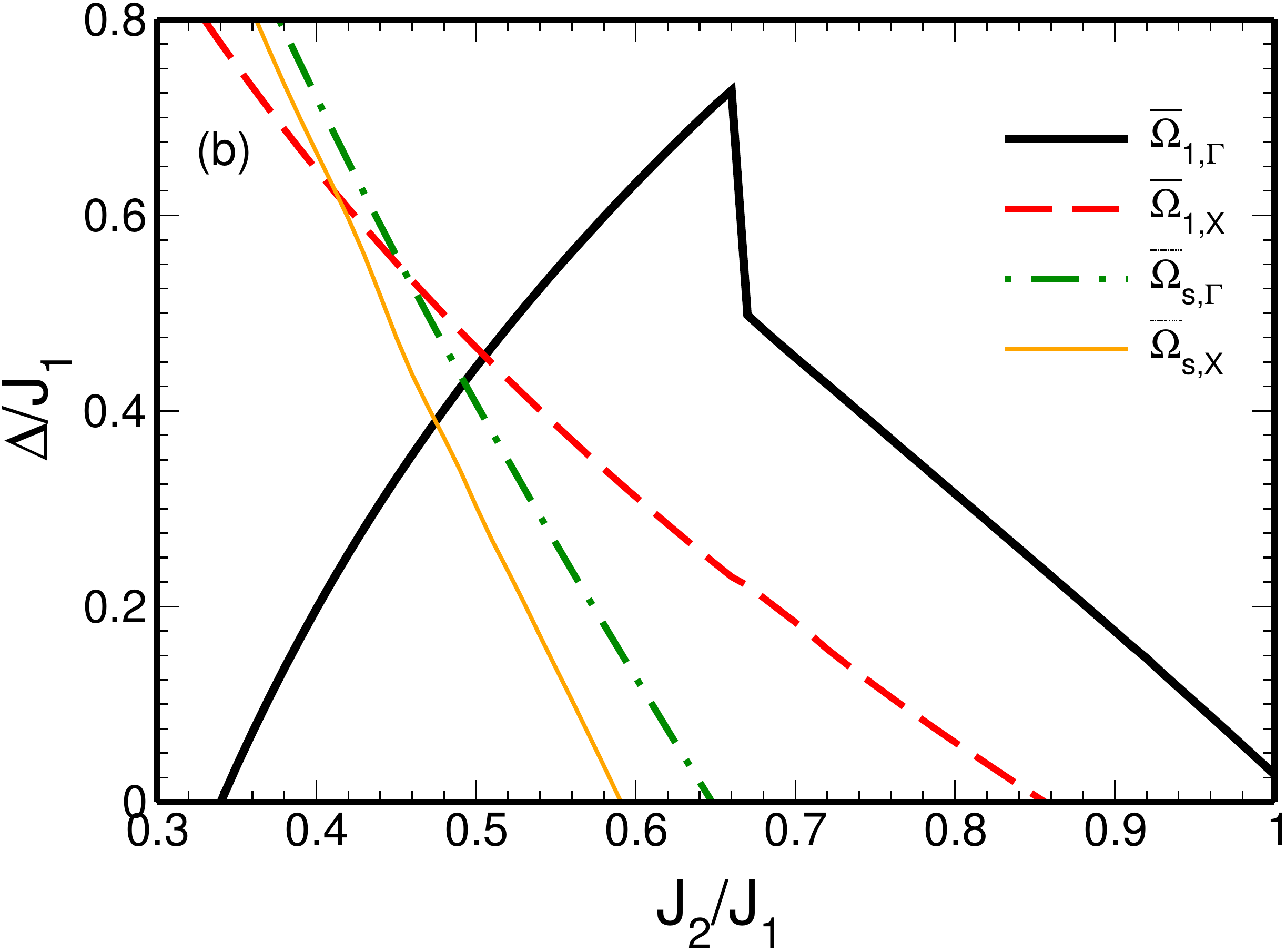}
}
\caption{(Color online) The excitation gap as a function of $J_2/J_1$
  at (a) the harmonic and (b) the cubic--quartic approximations. 
  $\Omega_{1,\Gamma}/\bar{\Omega}_{1,\Gamma}$ (thick solid black line) and
  $\Omega_{1,X}/\bar{\Omega}_{1,X}$ (dashed red line) are respectively the
  energy of the lowest--energy triplet excitation at the
  $\Gamma = (0,0)$ and ${\bf X} = (\pi/2,0)$ points, see
  Figs.~\ref{fig:disp} and \ref{fig:model}(c).
  $\Omega_{s,\Gamma}/\bar{\Omega}_{s,\Gamma}$ (dotted--dashed green line) and
  $\bar{\Omega}_{s,X}$ (thin solid orange line) are respectively the
  energy of the singlet excitations at the
  $\Gamma$ and ${\bf X}$ points.}
\label{fig:gap}
\end{figure*}

In order to diagonalize the triplet sector $\mathcal{H}_{20}$, 
it is useful to introduce the six--component vector
\[
  \Psi^\dagger_{\bk\alpha} = \left( t^\dagger_{1,\bk,\alpha}\;\;
        t^\dagger_{2,\bk,\alpha}\;\; t^\dagger_{3,\bk,\alpha}\;\;
        t_{1,-\bk,\alpha}\;\; t_{2,-\bk,\alpha}\;\; t_{3,-\bk,\alpha}\right)
\]
which allow us to rewrite Eq.~\eqref{ham-harmonic} in matrix form:
\begin{equation}
     \mathcal{H} = E'_0 + \mathcal{H}_{02}
       + \frac{1}{2}\sum_\bk \Psi^\dagger_{\bk\alpha} \hat{H}_\bk \Psi_{\bk\alpha}.
\label{ham-harmonic-matrix}
\end{equation}
Here
\[
  E'_0 = E_0 - \frac{3}{2}\sum_{a=1,2,3}\sum_\bk A^{aa}_\bk  
\]
and the $6\times 6$ matrix $\hat{H}_\bk$ reads
\begin{equation}
 \hat{H}_\bk = \left( \begin{array}{cc}
                          \hat{A}_\bk & \hat{B}_\bk \\
                          \hat{B}_\bk & \hat{A}_\bk
                      \end{array}  \right)
\label{matrix-AB}
\end{equation}
with $\hat{A}_\bk$ and $\hat{B}_\bk$ being $3\times 3$ Hermitian 
matrices whose elements are $A^{ab}_\bk$ and $B^{ab}_\bk$ respectively.
Although the diagonalization of the $6\times 6$ problem is quite involved 
(we briefly outline the analytical procedure in Appendix \ref{ap:diag}),
it is possible to show that, after the diagonalization, 
Eq.~\eqref{ham-harmonic-matrix} acquires the form
\begin{equation}
     \mathcal{H} = E_{EGS} + \mathcal{H}_{02}
      + \frac{1}{2}\sum_\bk \Phi^\dagger_{\bk\alpha} \hat{H}'_\bk \Phi_{\bk\alpha},
\label{ham-harmonic-matrix2}
\end{equation}
where 
\begin{equation}
  E_{EGS} = E_0 + \frac{3}{2}\sum_{a,\bk} \left( \Omega_{a,\bk} - A^{aa}_\bk \right)
\label{egs}
\end{equation}
is the ground state energy, the $6\times 6$ matrix $\hat{H}'_\bk$ reads
\[
 \hat{H}'_\bk = \left( \begin{array}{cc}
                          \hat{h}_\bk & 0 \\
                            0        & \hat{h}_\bk
                         \end{array}  \right)
\;\;\; {\rm with}\;\;\;
 \hat{h}_\bk = \left( \begin{array}{ccc}
                \Omega_{1,\bk}    &         0     &  0 \\
                          0     & \Omega_{2,\bk}  & 0 \\
                          0     &         0     & \Omega_{3,\bk} \\
                      \end{array}  \right),
\]
and the six--component vector $\Phi^\dagger_{\bk\alpha}$ is given by
\[
      \Phi^\dagger_{\bk\alpha} = \left( b^\dagger_{1,\bk,\alpha}\;\;
          b^\dagger_{2,\bk,\alpha}\;\; b^\dagger_{3,\bk,\alpha}\;\;
          b_{1,-\bk,\alpha}\;\; b_{2,-\bk,\alpha}\;\; b_{3,-\bk,\alpha}\right).
\]
The relation between the two set of boson operators
$t$ and $b$ is
\begin{equation}
 \Phi_{\bk\alpha} = \hat{M}_\bk\Psi_{\bk\alpha},
 \;\;\;\; {\rm where} \;\;\;\;
 \hat{M}_\bk = \left( \begin{array}{cc}
                            \hat{U}^\dagger_\bk & -\hat{V}^\dagger_\bk \\
                          -\hat{V}^\dagger_\bk & \hat{U}^\dagger_\bk
                            \end{array}  \right)
\label{bog-transf}
\end{equation}
with $\hat{U}_\bk$ and $\hat{V}_\bk$ being $3\times 3$ matrices
whose elements are the Bogoliubov coefficients 
$u^{ab}_\bk$ and $v^{ab}_\bk$. The explicitly expressions of the
triplet excitation energies $\Omega_{a,\bk}$ and the the Bogoliubov
coefficients $u^{ab}_\bk$ and $v^{ab}_\bk$ in terms of the 
$A^{ab}_\bk$ and $B^{ab}_\bk$ functions can be found in 
Appendix \ref{ap:diag}.

Finally, from the saddle points conditions $\partial E_0/\partial N_0
= 0$ and $\partial E_0/\partial \mu = 0$, self-consistent equations
for $\mu$ and $N_0$ follow, namely 
\begin{eqnarray}
 \mu &=& -2J_1 + \frac{1}{2}J_2 + \frac{3}{2N'}\sum_{a,\bk} \left[
                    \frac{\partial \Omega_{a,\bk}}{\partial N_0} -
                    \frac{1}{N_0}B^{aa}_\bk \right],  
\nonumber \\
&& \label{self-eq} \\
 N_0 &=& 1 + \frac{3}{2N'}\sum_{a,\bk} \left[ 1 +
                     \frac{\partial \Omega_{a,\bk}}{\partial\mu} \right]. 
\nonumber
\end{eqnarray}
Once $\mu$ and $N_0$ are numerically calculated, the triplet 
$\Omega_{a,\bk}$ and the singlet $\Omega_s = E_{s1} - \mu$ 
excitation energies are completely determined.

We numerically solve the self--consistent equations \eqref{self-eq}
and find solutions within the range $0.26 < J_2/J_1 < 1.0$
as indicated in Fig.~\ref{fig:phase-diag}(a).
The behaviour of the parameters $N_0$ and $\mu$ and the ground state 
energy \eqref{egs} as a function of $J_2/J_1$ are respectively 
shown in Figs.~\ref{fig:egs}(a) and (b). 
One sees that $N_0$ has a maximum at $J_2 = 0.58\,J_1$ and that $E_{EGS}$
monotonically increases with $J_2/J_1$. For comparison, we include the
ground state energy of the dimerized columnar [Fig.~\ref{fig:dimer}(a)]
and staggered [Fig.~\ref{fig:dimer}(b)] VBS phases
as obtained from the (dimer) bond--operator theory at the harmonic 
level (see Appendix \ref{ap:col-stag} for details). 
Note that the plaquette VBS state is the lowest--energy one and 
that it extends over a region of the parameter space much larger 
than the dimerized VBSs: the columnar VBS is stable for 
$0.38 < J_2/J_1 < 0.57$ while the staggered VBS 
only for $0.44 < J_2/J_1 < 0.56$.

Figures~\ref{fig:disp}(a) and (b) shows the energy of the triplet 
$\Omega_{a,\bk}$ (solid and dotted--dashed lines) and the singlet $\Omega_s$ (dashed line)
excitations for $J_2 = 0.48$ and $0.56\,J_1$, respectively. Recall that
$\Omega_s$ is dispersionless in the harmonic approximation.
One sees that for $J_2 = 0.48\,J_1$, the minimum (gap) of the triplet
dispersion relation occurs at the center of the tetramerized Brillouin
zone [$\Gamma$ point, see Fig.~\ref{fig:model}(c)] while for 
$J_2 = 0.56\,J_1$, at the $X$ point. As shown in Fig.~\ref{fig:gap}(a), 
such a changing in the momentum associated with the excitation gap takes place 
at $J_2 = 0.50\,J_1$. Interestingly, the gap changes from a 
triplet gap to a singlet one at $J_2 = 0.82J_1$.
Finally, note that the excitation gap is always finite
within the parameter region $0.26 < J_2/J_1 < 1.0$, i.e., there is 
no indication of a continuous quantum phase transition at any 
critical coupling $J_2$.

\begin{figure*}[t]
\centerline{\includegraphics[width=14.0cm]{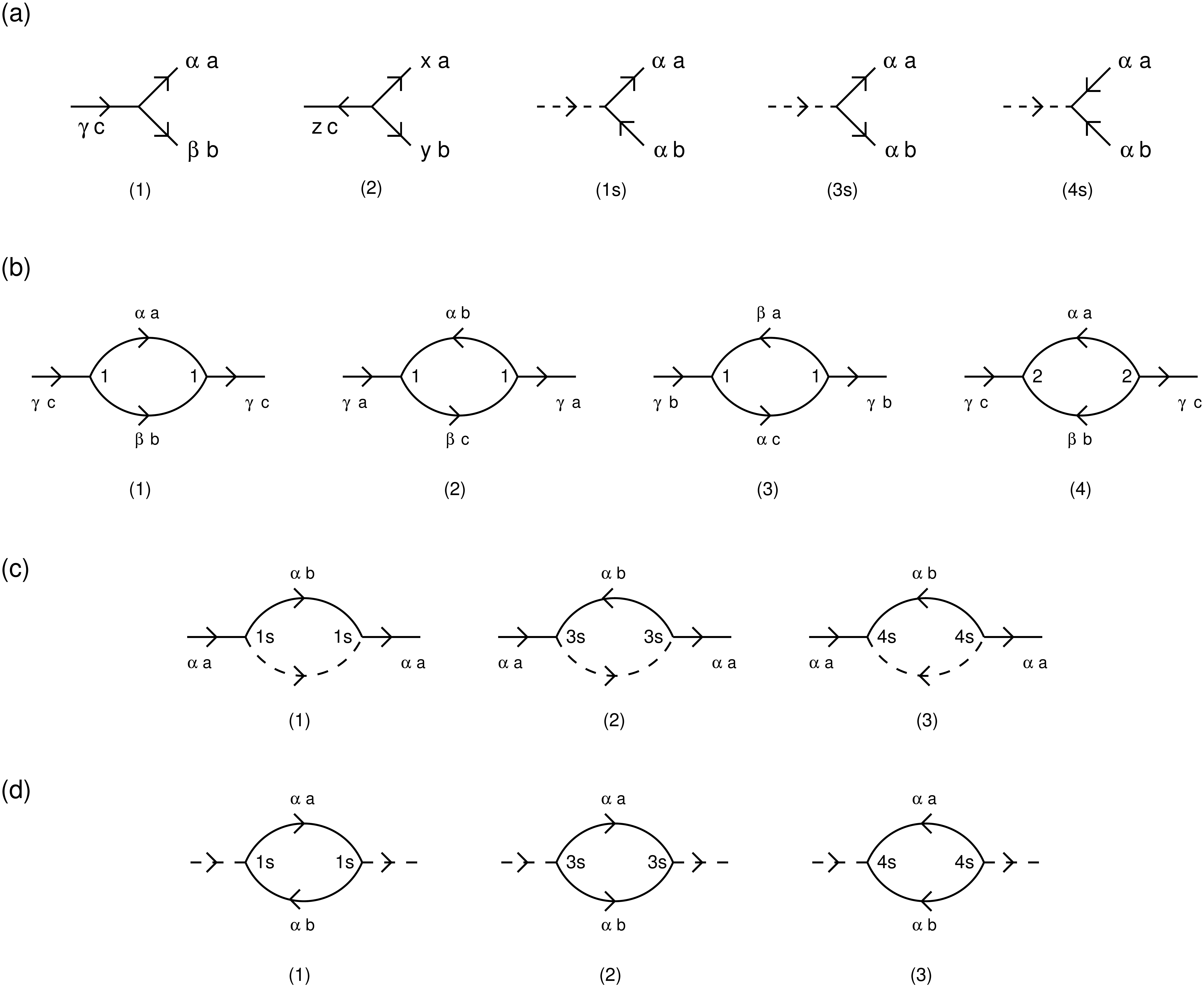}}
\caption{(a) Renormalized cubic vertices  
 $\Gamma^{abc}_{1,\bk,\bp}$, $\Gamma^{abc}_{2,\bk,\bp}$,
 $\Gamma^{ab}_{1s,\bk,\bp}$, $\Gamma^{ab}_{3s,\bk,\bp}$, and
 $\Gamma^{ab}_{4s,\bk,\bp}$, with
 $(\alpha,\beta,\gamma) = (x,y,z),\, (z,x,y),\, (y,z,x)$
 and $a,b,c = 1,2,3$.
 (b) and (c) Lowest--order diagrams derived from the renormalized
 cubic vertices $(1)$, $(2)$, $(1s)$, $(3s)$, and $(4s)$
 that contributes to the normal triplet self--energies 
 $\Sigma_a(\bk,\omega)$. 
 (d) Similar to the singlet self--energy $\Sigma_s(\bk,\omega)$.
 In each diagram, the solid and dashed lines correspond respectively
 to the bare (harmonic) $b$ triplet [Eq.~\eqref{bare-green-triplet}] 
 and $s_1$ singlet [Eq.~\eqref{bare-green-singlet}] propagators.
}
\label{fig:vertices}
\end{figure*}

\section{Cubic--quartic approximation}
\label{sec:cubic}

Since the energy of the singlet $s_1$ and the triplet $t_1$ excitations
are quite close for $J_2 \approx 0.5\, J_1$ [see Figs.~\ref{fig:disp}(a) 
and (b)], it is interesting to consider the effects of the cubic 
interaction $\mathcal{H}_{21}$ [Eq.~\eqref{h20}]. Moreover, we have 
recently shown that cubic (triplet--triplet--triplet) interactions 
provide important renormalizations to the harmonic (mean--field) 
excitation spectrum of a dimerized VBS phase 
in a frustrated quantum magnet.\cite{doretto12} 
Motivated by these two points, in this section we consider both cubic 
terms $\mathcal{H}_{30}$ and $\mathcal{H}_{21}$ within second--order
perturbation theory and calculate the corrections to the harmonic 
results determined in the previous section.
We also consider the quartic terms $\mathcal{H}_{40}$ [Eq.~\eqref{h40}]
and $\mathcal{H}_{22}$ [Eq.~\eqref{h22}] within the (no self-consistent) 
Hartree--Fock approximation. Although the quartic terms   
provide very small corrections to the harmonic results, they are
important in the determination of the critical couplings.
Such a procedure constitutes the so--called 
cubic--quartic approximation.

The first step is to express $\mathcal{H}_{30}$ and $\mathcal{H}_{21}$ 
in terms of the bosons $b$. With the help of Eq.~\eqref{bog-transf}, 
it is possible to show that
\begin{eqnarray}
\mathcal{H}_{30} &=& \frac{1}{\sqrt{N'}}\sum_{\bk,\bp} \left[
                   \sideset{}{'}\sum_{\alpha,\beta,\gamma}
                   \Gamma^{abc}_{1,\bk,\bp}
              \left (b^\dagger_{a,\bk-\bp, \alpha}b^\dagger_{b,\bp,\beta}b_{c,\bk,\gamma}
                       + {\rm H.c.} \right) \right.
\nonumber \\
&& \nonumber \left. \right. \\
             &+& \left.  \Gamma^{abc}_{2,\bk,\bp}
                \left( b^\dagger_{a,\bk-\bp,x}b^\dagger_{b,\bp ,y}b^\dagger_{c,-\bk, z}
                 + {\rm H.c.} \right) \right]
\label{h30-v2}
\end{eqnarray}
and
\begin{eqnarray}
\mathcal{H}_{21} &=& \frac{1}{\sqrt{N'}}\sum_{\bk,\bp}
                   \left[ \Gamma^{ab}_{1s,\bk,\bp}
                    b^\dagger_{a,\bk-\bp, \alpha}b_{b,-\bp,\alpha}s_{1,\bk}
                       + {\rm H.c.} \right. 
\nonumber \\
&& \left. \right. \nonumber \\
             &+& \left. \Gamma^{ab}_{3s,\bk,\bp}
                  b^\dagger_{a,\bk-\bp,\alpha}b^\dagger_{b,\bp ,\alpha}s_{1,\bk}
                 + {\rm H.c.} \right.
\nonumber \\
&& \left. \right. \nonumber \\
             &+& \left. \Gamma^{ab}_{4s,\bk,\bp}
                  b_{a,-\bk+\bp,\alpha}b_{b,-\bp ,\alpha}s_{1,\bk}
                 + {\rm H.c.} \right].
\label{h21-v2}
\end{eqnarray}
Here $a,b,c = 1,2,3$ (summation over repeated indices
is assumed), the sum over $\alpha,\, \beta, \, \gamma$ has only three
terms, $(\alpha,\beta,\gamma) = (x,y,z),\, (z,x,y),\, (y,z,x)$, and
the expressions of the renormalized cubic vertices  
$\Gamma^{abc}_{1/2,\bk,\bp}$ and $\Gamma^{ab}_{1s/3s/4s,\bk,\bp} $ 
[see Fig.~\ref{fig:vertices}(a)] 
in terms of the Bogoliubov coefficients $u^{ab}_\bk$ and $v^{ab}_\bk$
are given in Appendix \ref{ap:cubic}. 

\begin{figure*}[t]
\centerline{\includegraphics[width=6.5cm]{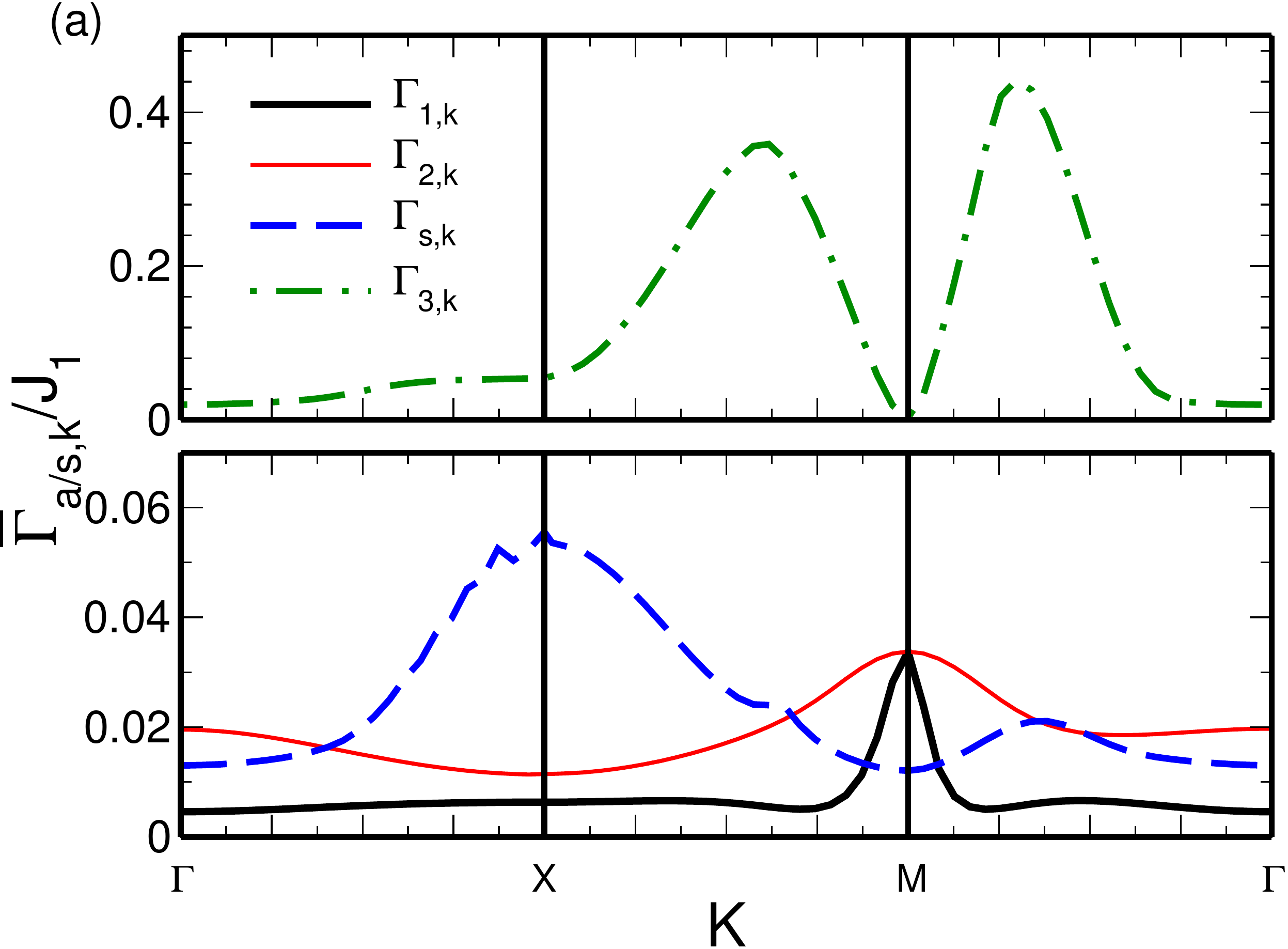}
            \hskip2.0cm
            \includegraphics[width=6.5cm]{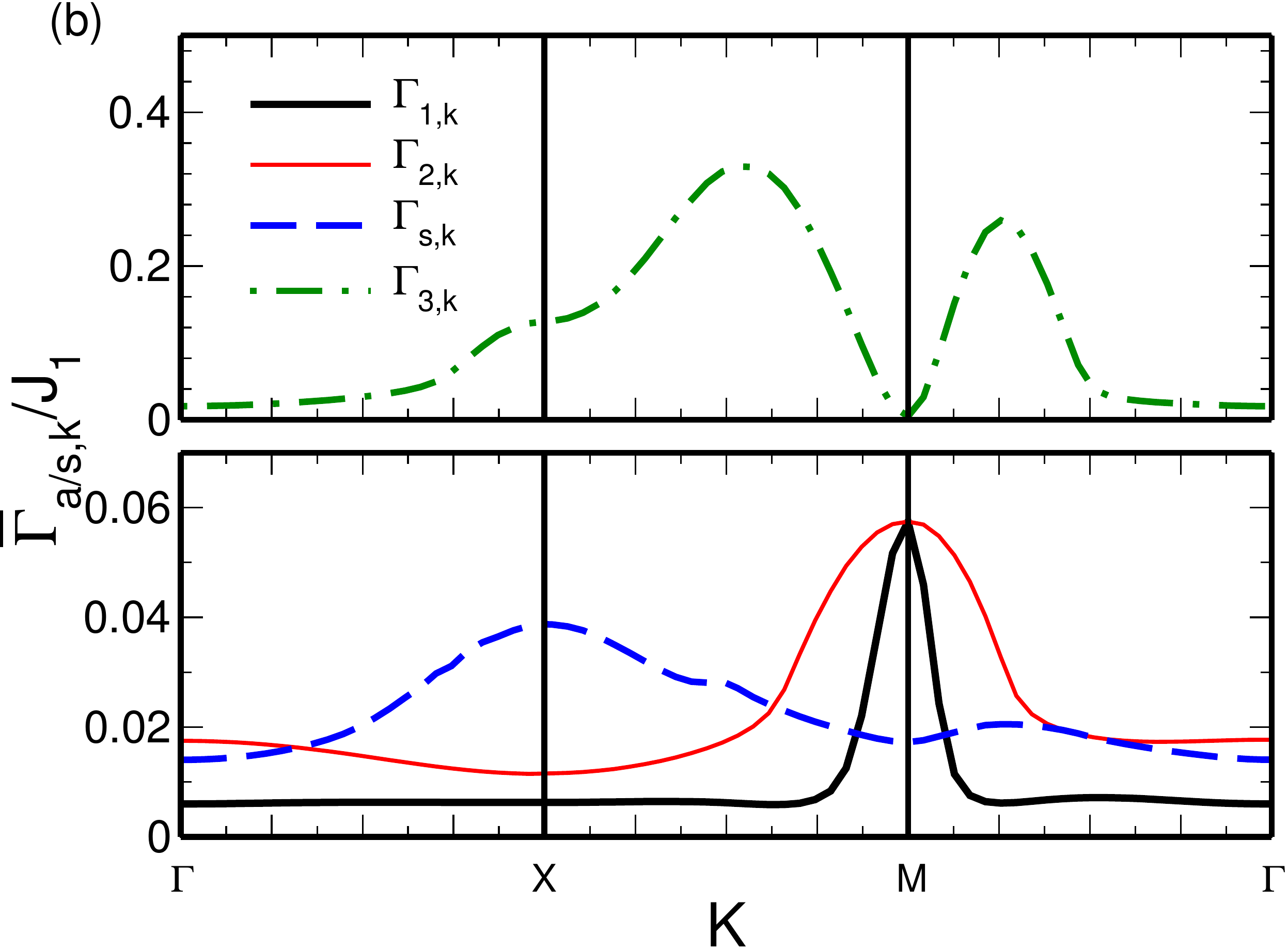}
}
\caption{(Color online) Decay rate (cubic--quartic approximation) of the 
          singlet $\bar{\Gamma}_{s,\bk}$ (dashed blue line) and the 
          triplet $\bar{\Gamma}_{1,\bk}$ (thick solid black line), 
                  $\bar{\Gamma}_{2,\bk}$ (thin solid red line), and  
                  $\bar{\Gamma}_{3,\bk}$ (dotted--dashed green line) 
          excitations along paths in the tetramerized
          Brillouin zone [Fig.~\ref{fig:model}(c)] 
          for (a) $J_2 = 0.48\,J_1$ and (b) $J_2 = 0.56\,J_1$.}
\label{fig:decay}
\end{figure*}

\begin{figure}[b]
\centerline{\includegraphics[width=6.5cm]{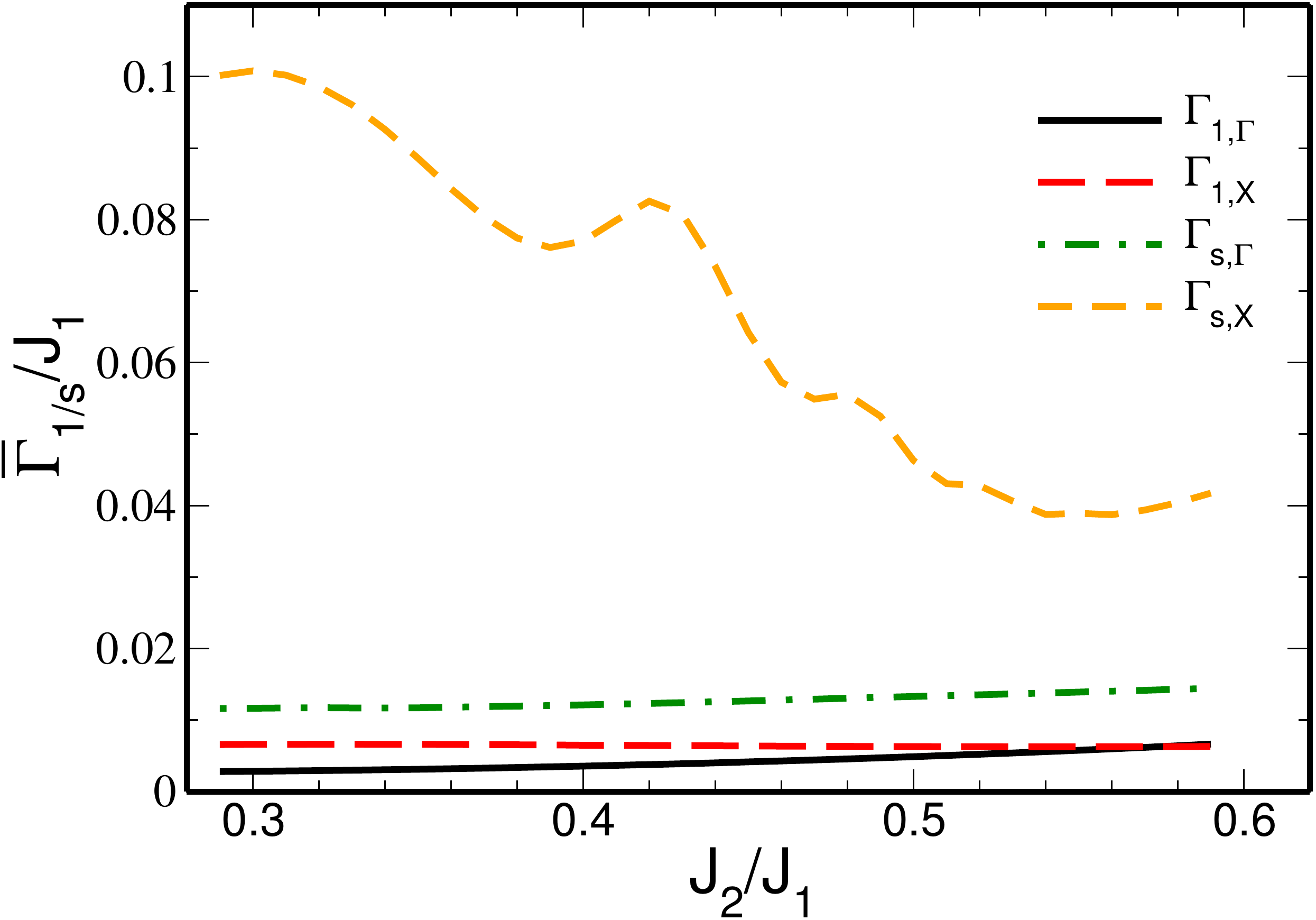}
}
\caption{(Color online) Decay rates as a function of $J_2/J_1$
         (cubic--quartic approximation). 
         $\bar{\Gamma}_{1,\Gamma}$ (solid black line) and
         $\bar{\Gamma}_{1,X}$ (dashed red line) are respectively the
         decay rates of the lowest--energy triplet excitation at the
         $\Gamma = (0,0)$ and ${\bf X} = (\pi/2,0)$ points, see
         Figs.~\ref{fig:decay} and \ref{fig:model}(c).
         $\bar{\Gamma}_{s,\Gamma}$ (dotted--dashed green line) and
         $\bar{\Gamma}_{s,X}$ (dashed orange line) are respectively the
         decay rates of the singlet excitation at the
         $\Gamma$ and ${\bf X}$ points.
}
\label{fig:decay2}
\end{figure}

Figures~\ref{fig:vertices}(b) and (c) show the lowest--order diagrams
that contribute to the (normal) triplet self--energies
$\Sigma_a(\bk,\omega)$ with $a = 1,2,3$, while Fig.~\ref{fig:vertices}(d) 
shows the ones related to the singlet self--energy
$\Sigma_s(\bk,\omega)$. The solid line in each diagram
corresponds to the bare (harmonic) $b$ triplet propagator,
\begin{equation}
  G_{0,a}^{-1}(\bk,\omega) = \omega - \Omega_{a,\bk} + i\delta,
\label{bare-green-triplet}
\end{equation}
and the dashed line denotes the bare $s_1$ singlet propagator,
\begin{equation}
  G_{0,s}^{-1}(\omega) = \omega - \Omega_s + i\delta,
\label{bare-green-singlet}
\end{equation}
with $\Omega_s = E_{s1} - \mu$. Hereafter, we omit the $\alpha$ index
in the triplet Green's functions and self--energies since the 
$x,\,y,\,z$  triplet branches for each $a$ are degenerate.
Note that there are no bare anomalous $b$ propagators. Although they
can be generated in perturbation theory, we neglect them in the 
following (for details, see note 40 from Ref.~\onlinecite{doretto12}).
Using standard diagrammatic techniques for bosons at zero temperature, 
we find that only the diagrams $(b1)$, $(b4)$, $(c1)$, $(c3)$,
$(d2)$, and $(d3)$ shown in Fig.~\ref{fig:vertices} are finite, and  
therefore, 
\begin{eqnarray}
  \Sigma_a(\bk,\omega) &=&  \Sigma^{(b1)}_a(\bk,\omega) 
                       + \Sigma^{(b4)}_a(\bk,\omega) 
\nonumber \\
&& \nonumber \\
  &+& \Sigma^{(c1)}_a(\bk,\omega)
     + \Sigma^{(c3)}_a(\bk,\omega),
\label{self-triplet}
\end{eqnarray}
and
\begin{equation}
  \Sigma_s(\bk,\omega) =  \Sigma^{(d2)}_s(\bk,\omega)
                       +  \Sigma^{(d3)}_s(\bk,\omega).
\label{self-singlet}
\end{equation}
The expressions of the different components of the
self--energies \eqref{self-triplet} and \eqref{self-singlet} 
are shown in Appendix \ref{ap:cubic}.

Turning to the quartic terms $\mathcal{H}_{40}$ 
and $\mathcal{H}_{22}$, it is possible to show that
\begin{eqnarray}
\mathcal{H}_{40} &=& E^{(4)}_{EGS} + \sum_{a,b,\alpha} \sum_\bp  
                    A^{HF}_{ab,\bp}\;b^\dagger_{a,\bp,\alpha}b_{b,\bp,\alpha}
\nonumber \\
&& \nonumber \\
               &+&  \left[ B^{HF}_{ab,\bp}\;b^\dagger_{a,\bp,\alpha}b^\dagger_{b,-\bp,\alpha}
                    + {\rm H.c.} \right] + \mathcal{O}(b^4) 
\label{h40-v2}
\end{eqnarray}
and
\begin{eqnarray}
\mathcal{H}_{22} &\approx& \sum_\bp A^{HF}_{s,\bp}\; s^\dagger_{1,\bp}s_{1,\bp}
                   + \frac{1}{2}\left( B^{HF}_{s,\bp}\;
                      s^\dagger_{1,-\bp}s^\dagger_{1,\bp} + {\rm H.c.} \right),
\nonumber \\
\label{h22-v2}
\end{eqnarray}
where the constant $E^{(4)}_{EGS}$ and the coefficients $ A^{HF}_{ab/s,\bp}$ and
$ B^{HF}_{ab/s,\bp}$ can be found in Appendix \ref{ap:cubic}.
Note that Eq.~\eqref{h40-v2} is not diagonal in the $a$ and $b$ indices.
The (normal) triplet and singlet self--energies are then respectively 
given by
\begin{equation}
  \Sigma^{HF}_{ab}(\bk) = A^{HF}_{ab,\bp},
  \;\;\;\; {\rm and} \;\;\;\;
  \Sigma^{HF}_s(\bk) = A^{HF}_{s,\bp}.
\label{self-quartic}
\end{equation}

The renormalized singlet $\bar{\Omega}_{s,\bk}$ and triplet
$\bar{\Omega}_{a,\bk}$ excitation energies and the decay
rates $\bar{\Gamma}_{s/a,\bk}$ are given by the poles of
the corresponding Green's function $G_{s/a}(\bk,\omega)$:
\[
  G^{-1}_{s/a}(\bk,\omega) = \omega - \Omega_{s/a,\bk} -
  \Sigma_{s/a}(\bk,\omega) - \Sigma^{HF}_{s/aa}(\bk) = 0.
\]
Note that in addition to the anomalous Hartree-Fock self--energies,
the normal ones with $a\not= b$ are also neglected,
since it significantly simplifies the determination of the poles
of the Green's function. 
The above equation is solved within the on-shell
approximation,\cite{doretto12,zhito09} where the self--energy is
evaluated at the bare (harmonic) single--particle energy:
\[
   \bar{\Omega}_{s/a,\bk} - i\bar{\Gamma}_{s/a,\bk} - \Omega_{s/a,\bk}
     - \Sigma_{s/a}(\bk,\Omega_{s/a,\bk}) -\Sigma^{HF}_{s/aa}(\bk) = 0.
\] 
Such a procedure, which is less involved than the off--shell 
approximation adopted in Ref.~\onlinecite{doretto12}, 
provides reasonable results for the excitation spectra 
(see below) without the discontinuities and logarithmic 
singularities reported in Ref.~\onlinecite{zhito09}.

Finally, the ground state energy reads
\begin{equation}
 \bar{E}_{EGS} = E_{EGS} +  E^{(3)}_{EGS} + E^{(4)}_{EGS},
\label{total-egs}
\end{equation}
where $E_{EGS}$ is the harmonic term \eqref{egs} and the 
expressions of the corrections due to cubic
[$E^{(3)}_{EGS}$] and quartic [$E^{(4)}_{EGS}$] interactions are 
presented in Appendix~\ref{ap:cubic}.

The renormalized singlet $\bar{\Omega}_{s,\bk} $ and triplet 
$\bar{\Omega}_{a,\bk}$ excitation spectra for $J_2 = 0.48$ and $0.56\,J_1$
are respectively shown in Figs.~\ref{fig:disp}(c) and (d)
while the corresponding decay rates $\bar{\Gamma}_{s/a,\bk}$,
in Figs.~\ref{fig:decay}(a) and (b). 
One sees that the excitation energies decrease as compared to the 
harmonic ones, an effect similar to what we have found in the
triangular lattice quantum magnet.\cite{doretto12} 
In particular, the singlet excitation branch, which now acquires a 
dispersion, is the lowest--energy excitation for both configurations. 
We find that the contributions of $\Sigma^{(b1)}_a(\bk,\omega)$ and 
$\Sigma^{(b4)}_a(\bk,\omega)$ to the renormalized triplet spectra
$\bar{\Omega}_{a,\bk}$ are much larger than the ones associated with 
$\Sigma^{(c1)}_a(\bk,\omega)$ and $\Sigma^{(c3)}_a(\bk,\omega)$.
Moreover, we also find that the renormalizations due to the 
cubic vertices are stronger than the ones associated with the
quartic interactions. The most important contributions 
of $\Sigma^{HF}_{aa}(\bk,\omega)$ to the triplet excitation spectra
occurs around $J_2 = 0.30\,J_1$.

The behaviour of the excitation gap as a function of $J_2/J_1$ is shown in
Fig.~\ref{fig:gap}(b). Note that the gap vanishes at the critical 
couplings $J_2 = 0.34$ and $0.59\,J_1$, indicating that the plaquette 
VBS phase is stable only within the parameter region 
$0.34 < J_2/J_1 < 0.59$. Such a result sharply contrasts with the
ones obtained within the harmonic approximation 
[see Figs.~\ref{fig:phase-diag}(a) and (b)]. 
Moreover, as $J_2/J_1$ increases, the excitation gap changes from
a triplet gap to a singlet one:
for $0.34 < J_2/J_1 < 0.48$, the gap is
associated with a singlet--triplet excitation at the $\Gamma$ point 
while, for  $0.48 < J_2/J_1 < 0.59$, with a singlet--singlet 
excitation at the ${\bf X} = (\pi/2,0)$ point. 
Recall that such a change in the nature of the excitation gap for
$J_2 \approx 0.5\,J_1$ is similar to the behaviour of the
$J_1$--$J_2$ model on a single plaquette 
[see Fig.~\ref{fig:plaquette}(b)].  

In addition to renormalize downward the excitation energies,
the cubic vertices may also enable two--particle decay of the 
singlet and triplet modes [Figs.~\ref{fig:decay}(a) and (b)].
In particular, note that for  $J_2 = 0.48$ and $0.56\,J_1$,
the triplet decay rate $\bar{\Gamma}_{1,\Gamma} \approx 0$  
while the singlet one $\bar{\Gamma}_{s,X}$ is finite.  
Indeed, while the former is constant, the latter has
an almost monotonic behaviour, decreasing with $J_2/J_1$,
see Fig.~\ref{fig:decay2}. 
Such a result indicates that the excitation gap 
acquires a finite decay rate for $J_2 > 0.48\,J_1$,
that decreases and (almost) vanishes closes to the critical 
coupling $J_2 = 0.59\,J_1$.\cite{comment2}

Finally, we should note that cubic and quartic vertices 
provide very small corrections to the harmonic ground state energy, 
see Fig.~\ref{fig:egs}(b).

\section{Discussion}
\label{sec:discussion}

According to the harmonic bond--operator theory (Sec.~\ref{sec:harmonic}),  
the plaquette VBS phase has lower energy than the 
dimerized columnar [Fig.~\ref{fig:dimer}(a)] and 
staggered [Fig.~\ref{fig:dimer}(b)] ones. Moreover, the ground state 
energy of the plaquette phase monotonically increases with $J_2/J_1$ 
while, for the dimerized phases, $E_{EGS}$ is a convex function with 
a minimum around $J_2 = 0.5\,J_1$. The behaviour of the 
plaquette ground state energy qualitatively agrees with exact 
diagonalization data, which show that $E_{EGS}$ monotonically increases with $J_2/J_1$, 
reaches a maximum around $J_2 = 0.6\,J_1$, and then 
decreases.\cite{dagotto89,schulz92,richter10}  
Such an agreement could be seen as a further indication that 
the plaquette phase might set in within the disordered region
of the $J_1$--$J_2$ model.  
A similar behaviour for the ground state energy is also observed
in coupled cluster,\cite{darradi08}
hierarchical mean--field,\cite{isaev09} and
tensor network states\cite{yu12} calculations.
 
As mentioned in the Introduction (Sec.~\ref{sec:intro}), 
Zhitomirsky and Ueda\cite{zhito96} studied the plaquette VBS 
phase of the $J_1$--$J_2$ model within the bond--operator theory 
at the harmonic level without including the high--energy singlet 
state $|s_1\rangle$. They found that the plaquette phase is stable
for $0.08 < J_2/J_1 < 0.80$ and that it has
lower energy than the dimerized columnar VBS. 
In particular, for $J_2 = 0.50\,J_1$, they found that the excitation gap 
$\Delta = 0.85\,J_1$ while the ground state energy $E_{EGS} = -0.466\,J_1$. 
Although the region of stability of the plaquette phase that we 
arrive at [see Fig.~\ref{fig:phase-diag}(a)] differs from their results,
both harmonic (mean--field) calculations show that the 
plaquette VBS phase extends over a region much larger than the 
$J_1$--$J_2$ model paramagnetic one 
($0.4 \lesssim J_2/J_1 \lesssim 0.6$, see Sec.~\ref{sec:intro}).
Our mean--field results are in reasonable agreement with 
Ref.~\onlinecite{zhito96}: Recall that we also find that   
the plaquette VBS state is more stable than the dimerized
columnar state [Fig.~\ref{fig:egs}(b)]. 
Moreover, for $J_2 = 0.50\,J_1$, the gap $\Delta = 0.89\,J_1$ 
and the ground state energy $E_{EGS} = -0.472\,J_1$. 

As described in Sec.~\ref{sec:cubic}, cubic and quartic
vertices strongly modify the harmonic singlet and triplet excitation 
spectra of the $J_1$--$J_2$ model, similar to what we have recently 
found for a triangular lattice AFM.\cite{doretto12} 
One important consequence is that the region of stability of the 
plaquette VBS phase ($0.34 < J_2/J_1 < 0.59$) is reduced 
as compare with the harmonic one [see Figs.~\ref{fig:phase-diag}(a) and (b)]
and it is now quite close to the disordered region of the 
$J_1$--$J_2$ model found in previous calculations, see Sec.~\ref{sec:intro}.
Such a result shows that cubic and quartic interactions are indeed
relevant for a proper description of the plaquette VBS phase within
the bond operator approach. We should note that although the cubic 
corrections to the harmonic results are much larger than the quartic 
ones, the latter has an important role in the determination of the lower 
critical coupling: including only the cubic vertices, we find that 
the region of stability of the plaquette phase is $0.29 < J_2/J_1 < 0.59$.    

Although the region of stability derived within the cubic--quartic 
approximation almost agrees with the paramagnetic region of the 
$J_1$--$J_2$ model, the lower critical coupling $J_2 = 0.34\,J_1$ 
is smaller than the ones reported in the literature, i.e, 
$J_2 \approx 0.40\,J_1$, see Sec.~\ref{sec:intro}. 
In particular, it is even smaller than the one derived within
linear spin--wave theory, $J_2 \approx 0.38\,J_1$, (corrections up to
second order in the $1/S$ expansion of the sublattice magnetization 
even increase the lower critical coupling, i.e., the region of 
stability of the N\'eel phase increases when $1/S$ corrections are 
added to the linear spin--wave results, see 
Ref.~\onlinecite{igarashi93} for details).  
Differently from spin--wave theory, where $1/S$ can be taken as
a small parameter,\cite{igarashi93,zhito09} the bond 
operator formalism lacks such a quantity (in principle, the 
density of excited triplets can be considered as a small parameter, 
see Ref.~\onlinecite{kotov98} for details) and therefore, it is
difficult to systematically determine corrections to the mean--field
results. We believe that the results derived here could be improved
once: 
(a) the full singlet and triplet propagators, instead of the 
bare ones, are employed in the calculation of the normal triplet 
[Eqs.~\eqref{self-triplet}] and singlet [\eqref{self-singlet}] 
self--energies; 
(b) the anomalous cubic and quartic self--energies are considered; 
(c) the influence of the quintet excitations are taken into account; 
and/or 
(d) an alternative treatment of the constraint \eqref{constraint}
is employed (see Sec.~II.C from Ref.~\onlinecite{doretto12} 
for details).    
However, it is difficult to say which one is the most relevant
contribution to the determination of the phase boundary.

The nature of the excitation gap of the plaquette VBS phase is also affected
by cubic and quartic vertices: for $J_2 < 0.48\,J_1$, we find a triplet gap 
while for $J_2 > 0.48\,J_1$, a singlet one.
It should be contrasted with the harmonic approximation: the 
gap changes from a triplet gap to a singlet one at $J_2 = 0.82\,J_1$. 
Interestingly, one of the first exact diagonalization data\cite{dagotto89} 
for the $J_1$--$J_2$ model indicates that the excitation gap is associated with 
a singlet--singlet excitation for $0.50 < J_2/J_1 < 0.60$. We should also 
note that: (i) The hierarchical mean--field approach\cite{isaev09,gerardo13}
also indicates that the excitation gap changes from a triplet to a singlet
one, but at $J_2 \simeq 0.57\,J_1$; 
(ii) The DMRG calculations recently reported in Ref.~\onlinecite{jiang12},
which find some evidences for a ${\rm Z}_2$ spin--liquid phase,
point to a singlet gap smaller than the triplet one within the 
whole disordered region. 
 
Cubic and quartic vertices also influence the nature of the phase transitions at 
small and large $J_2$. Recall that (Sec.~\ref{sec:cubic}) for 
$J_2 = 0.34\,J_1$, a triplet gap vanishes, indicating a continuous quantum 
phase transition either to an ordered phase or to a mixed phase\cite{comment3} 
(N\'eel phase with plaquette modulation). 
As discussed in the Introduction, the former scenario is in 
favor of the deconfined quantum criticality theory\cite{senthil04} for the 
N\'eel--VBS transition while the latter scenario is in agreement with the 
Landau--Ginzburg framework. 
On the other hand, for $J_2 = 0.59\,J_1$, a singlet gap vanishes which, in principle, 
points to a continuous quantum phase transition to a dimerized columnar 
VBS phase: note that a suitable linear combination of $|s_0\rangle$ and 
$|s_1\rangle$ [see Eq.~\eqref{s0-s1} and Fig.~\ref{fig:singlets}] yields 
a (columnar) dimer state. Here, a continuous transition to a mixed phase
(columnar VBS with plaquette modulation) should not be excluded either.\cite{comment3}
Such a result is in contradiction with previous ones (see
Sec.~\ref{sec:intro})  which indicate that a first--order quantum phase transition
takes place at $J_2 \approx 0.60\,J_1$ from a quantum paramagnetic
phase to a collinear (ordered) one (see discussion below). 

Finally, in order to check the accuracy of our results, it is 
interesting to compare the ground state energy and the excitation gap
for $J_2 = 0.5\,J_1$, which is deep in the disordered phase, 
with the available data. Within the cubic--quartic approximation (Sec.~\ref{sec:cubic}), 
we find that $E_{EGS} = -0.477\,J_1$, which is in reasonable agreement with 
(plaquette) series expansion results,\cite{singh99}  
coupled cluster calculations,\cite{darradi08}  
the latest exact diagonalization data for $N = 40$ sites,\cite{richter10} 
and a very recent DMRG (Ref.~\onlinecite{gong13-2}, see also note at the 
end of Sec.~\ref{sec:summary}) that respectively indicate that
$E_{EGS} \simeq -0.485$, $-0.50$, $-0.499$, and $-0.497\,J_1$.
In this case, one notices that the different methods agree fairly well.
On the other hand, there is no consensus about the value of the excitation 
gap. For instances, we find for the triplet excitation gap $\Delta \simeq 0.018$, 
$0.12$, $0.30$, and $0.35\,J_1$ respectively derived within DMRG,\cite{gong13-2} 
Green function Monte Carlo,\cite{capriotti00}
hierarchical mean--field [see also note (i) above],\cite{isaev09} 
and exact diagonalization\cite{richter10}
approaches. Recall that [Fig.~\ref{fig:gap}(b)] we arrive at  
$\Delta = 0.30$ and $0.46\,J_1$ for the singlet and triplet excitation gaps, 
respectively, which are larger than the values reported in the literature. 
As discussed above, such excitations gaps could decrease if, for instance, 
cubic and quartic interactions are self-consistently considered.

\subsection{Consequences for the $J_1$--$J_2$ model}

The results that we have derived within the bond--operator theory
(cubic--quartic approximation) allow us to state that if a plaquette 
VBS phase sets in for $J_2 \approx 0.5\,J_1$, then such a phase displays a {\sl singlet}
excitation gap. This is the same feature of a possible
spin liquid phase described by recent DMRG simulations.\cite{jiang12}   
Therefore, the determination of the nature of the excitation gap is
not enough to make a distinction between the plaquette VBS phase and 
a ${\rm Z}_2$ spin--liquid for $J_2 \approx 0.5\,J_1$.

The fact that a singlet gap vanishes at $J_2 = 0.59\,J_1$ 
disagrees with previous calculations (see Sec.~\ref{sec:intro}). Such a result could indicate
that: 
(a) the plaquette--columnar VBS transition is indeed a true
quantum phase transition and a first--order columnar VBS--collinear
quantum phase transition takes place at a larger $J_2$,
(b) a first--order quantum phase transition to the collinear phase
pre--emptes the plaquette--columnar VBS transition, or
(c) a mixed columnar--plaquette phase\cite{ralko09} may set in within 
the disordered region. 
It should be mentioned that the possibility of a series of intermediate 
paramagnetic phases between the N\'eel and the collinear phases [scenario (a)] 
is discussed in Ref.~\onlinecite{sushkov01} and that the plaquette--columnar
quantum phase transition was studied by Kotov {\it et al.},\cite{kotov01}
who showed that such a quantum critical point belongs to the  
$O(1)$ universality class (equivalent to 3D Ising).

We intend to investigate the above scenario (c) 
within the bond--operator theory in a future publication.

\section{Summary}
\label{sec:summary}

In this paper, we revisited the work of Zhitomirsky and Ueda\cite{zhito96} 
and studied the plaquette VBS phase of the square lattice $J_1$--$J_2$ AFM 
model within the (tetramerized) bond--operator theory. 
We improved the previous analysis by including the high--energy singlet 
state within the description and perturbatively taking into account the 
effects of cubic (singlet--triplet--triplet and triplet--triplet--triplet) 
and quartic vertices above the harmonic (mean--field) results. 
We showed that cubic and quartic
interactions play an important role in the determination of the singlet and 
the triplet excitation spectra. As a consequence, the region of stability
of the plaquette phase is smaller than the harmonic one.   
Interesting, we found that at $J_2 = 0.48\,J_1$,
the excitation gap of the plaquette VBS phase changes from a triplet gap
to a singlet one, which vanishes at $J_2 = 0.59\,J_1$.

We would like in the near future to apply the formalism discussed 
here to study the stability of the plaquette VBS phase in some extensions
of the $J_1$--$J_2$ model. For instance, the square lattice $J_1$--$J_2$--$J_3$ 
AFM model, where there are some evidences\cite{mambrini06,reuther11} that 
the inclusion of a next--next--nearest--neighbor AFM coupling $J_3$ favors 
the stability of the plaquette phase. We also believe that effects of
anisotropy in the plaquette VBS phase can also be addressed. In this case,
one candidate is the square lattice $J^{XXZ}_1$--$J^{XXZ}_2$ AFM model recently 
considered in Ref.~\onlinecite{bishop08}.

As a final remark, we would like to mention that it would also be interesting
to consider the AFM $J_1$--$J_2$ model on the honeycomb lattice within the
procedure developed here. There are numerical evidences that a plaquette VBS 
phase may set in within the zero temperature phase diagram not only in the 
$J_1$--$J_2$ model\cite{ganesh13,zhu13,gong13} but also in  
the $J_1$--$J_2$--$J_3$ model.\cite{albuquerque11,bishop12}  
However, this is a much more involved task since the Hilbert space of   
six spins $S=1/2$ on a hexagon has 64 states: five singlet, 27 triplet,
25 quintet, and seven septet states. In this case, it is very difficult to determine
the bond operator representation, i.e., the equivalent of Eq.~\eqref{spin-bondop},
for the spin operators. 

{\sl Note added}. We recently became aware of DMRG calculations\cite{gong13-2}
which indicates that the plaquette VBS phase is stable for $0.50 < J_2/J_1 < 0.61$.
The authors also found that the N\'eel order vanishes for $J_2 > 0.44\,J_1$ and
that a possible gapless spin liquid phase may set in for $0.44 < J_2/J_1 < 0.50$.

\acknowledgments

We thank M. Vojta, E. Miranda, and A. O. Caldeira for helpful discussions and 
FAPESP, project No.~2010/00479-6, for the financial support.

\appendix

\section{Single--plaquette Hilbert space}
\label{ap:hilbert}

\begin{figure}[t]
\centerline{\includegraphics[width=5.5cm]{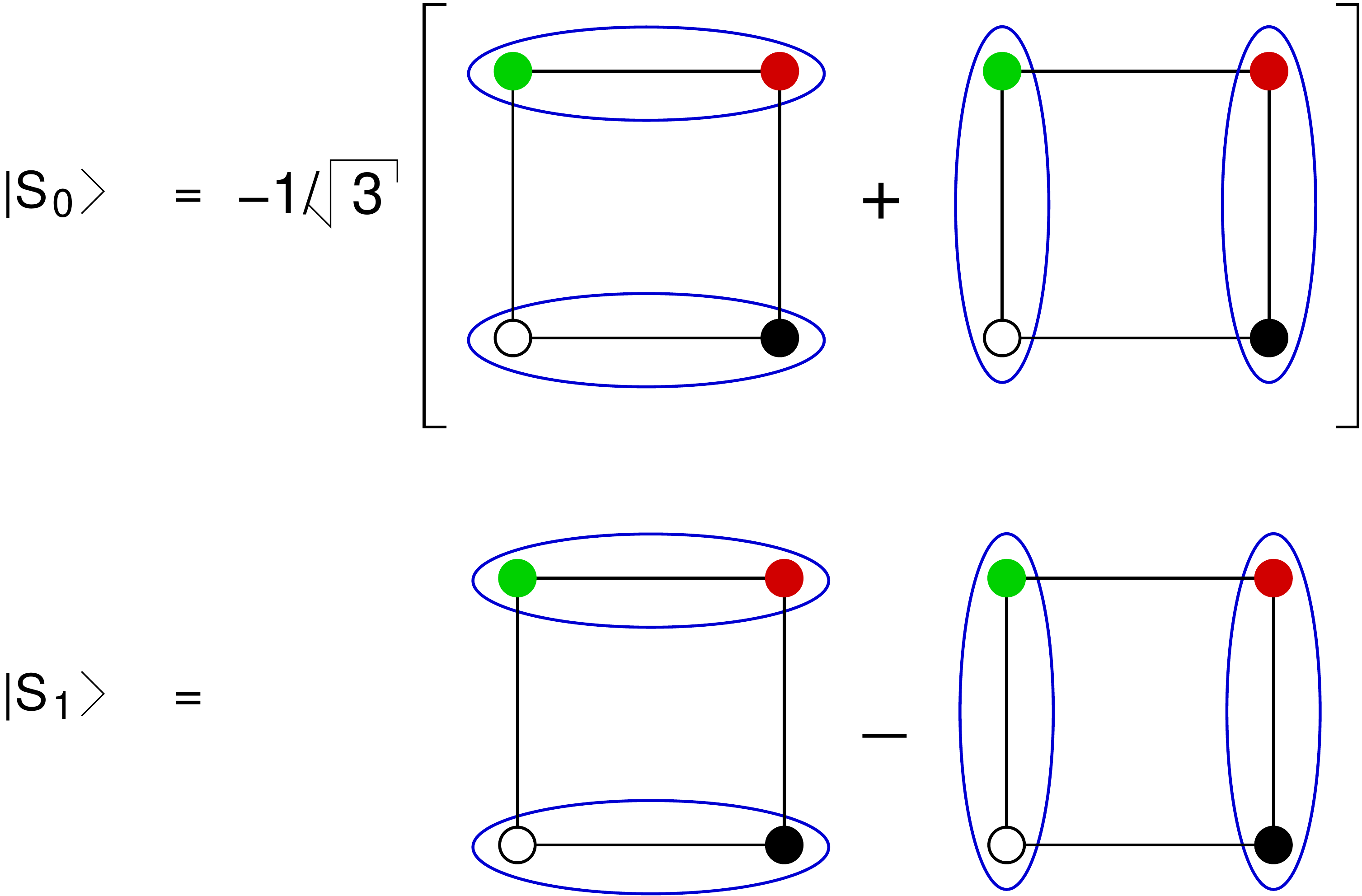}}
\caption{(Color online) Schematic representation of the singlet
         states $|s_0\rangle$ and $|s_1\rangle$ [Eq.~\eqref{s0-s1}].
         The symbols are the same as in Figs.~\ref{fig:model} and
         \ref{fig:dimer}.}
\label{fig:singlets}
\end{figure}

In this section, we provide the expansion of the eigenvectors of the Hamiltonian  
\eqref{ham-j1j2-plaq} (the two singlet, $|s_0\rangle$ and $|s_1\rangle$,
nine triplet, $| t_{a,\alpha} \rangle$ with $a=1,2,3$ and $\alpha = x,y,z$, 
and five quintet, $|d_0\rangle$, $|d_2\rangle$, and $|d_\alpha\rangle$, states)
in terms of the 16 states 
$|\uparrow\,\uparrow\,\uparrow\,\uparrow \rangle$,
$|\downarrow\,\uparrow\,\uparrow\,\uparrow \rangle$, 
$|\uparrow\,\downarrow\,\uparrow\,\uparrow \rangle$, 
$\ldots$, etc. It is possible to show that
\begin{eqnarray}
| s_0 \rangle   &=& \frac{1}{2\sqrt{3}}\left(
     2|\uparrow\, \downarrow\, \uparrow\, \downarrow \rangle 
   + 2|\downarrow\, \uparrow\, \downarrow\, \uparrow \rangle
    - |\uparrow\, \uparrow\, \downarrow\, \downarrow \rangle 
    - |\downarrow\, \downarrow\, \uparrow\, \uparrow \rangle \right. 
\nonumber \\
&& \nonumber \\
&& \left. \;\;\;\;\;\;
    - |\downarrow\, \uparrow\, \uparrow\, \downarrow \rangle 
    - |\uparrow\, \downarrow\, \downarrow\, \uparrow \rangle   \right), 
\nonumber\\
&& \nonumber \\
| s_1 \rangle   &=& \frac{1}{2}\left(
      |\downarrow\, \uparrow\, \uparrow\, \downarrow \rangle 
    + |\uparrow\, \downarrow\, \downarrow\, \uparrow \rangle
    - |\uparrow\, \uparrow\, \downarrow\, \downarrow \rangle 
    - |\downarrow\, \downarrow\, \uparrow\, \uparrow \rangle  \right),
\nonumber \\
&& \nonumber \\ 
| t_{1,\alpha} \rangle   &=& \frac{\lambda_\alpha}{2}\left(
      |\uparrow\, \uparrow\, \downarrow\, \uparrow \rangle 
    - |\downarrow\, \uparrow\, \uparrow\, \uparrow \rangle
  \mp |\uparrow\, \downarrow\, \downarrow\, \downarrow \rangle 
  \pm |\downarrow\, \downarrow\, \uparrow\, \downarrow \rangle  \right),
\nonumber \\
&& \nonumber \\
| t_{1,z} \rangle   &=& \frac{1}{2}\left(
      |\downarrow\, \downarrow\, \uparrow\, \uparrow \rangle 
    - |\uparrow\, \uparrow\, \downarrow\, \downarrow \rangle
    - |\uparrow\, \downarrow\, \downarrow\, \uparrow \rangle 
    + |\downarrow\, \uparrow\, \uparrow\, \downarrow \rangle   \right),
\nonumber \\
&& \nonumber \\
| t_{2,\alpha} \rangle   &=& \frac{\lambda_\alpha}{2}\left(
      |\uparrow\, \uparrow\, \uparrow\, \downarrow\, \rangle 
    - |\uparrow\, \downarrow\, \uparrow\, \uparrow \rangle
  \mp |\downarrow\, \uparrow\, \downarrow\, \downarrow \rangle 
  \pm |\downarrow\, \downarrow\, \downarrow\, \uparrow \rangle  \right),
\nonumber \\
&& \nonumber \\
| t_{2,z} \rangle   &=& \frac{1}{2}\left(
      |\downarrow\, \downarrow\, \uparrow\, \uparrow \rangle 
    - |\uparrow\, \uparrow\, \downarrow\, \downarrow \rangle
    + |\uparrow\, \downarrow\, \downarrow\, \uparrow \rangle 
    - |\downarrow\, \uparrow\, \uparrow\, \downarrow \rangle  \right),
\nonumber \\
&& \nonumber \\
| t_{3,\alpha} \rangle   &=& \frac{\lambda_\alpha}{2\sqrt{2}}\left(
      |\downarrow\, \uparrow\, \uparrow\, \uparrow \rangle 
    - |\uparrow\, \downarrow\, \uparrow\, \uparrow \rangle
    + |\uparrow\, \uparrow\, \downarrow\, \uparrow \rangle 
    - |\uparrow\, \uparrow\, \uparrow\, \downarrow \rangle \right. 
\nonumber \\
&& \left. \right. \nonumber \\
&& \left. \;\;\;\;\;\;
  \pm |\uparrow\, \downarrow\, \downarrow\, \downarrow \rangle 
  \mp |\downarrow\, \uparrow\, \downarrow\, \downarrow \rangle 
  \pm |\downarrow\, \downarrow\, \uparrow\, \downarrow \rangle 
  \mp |\downarrow\, \downarrow\, \downarrow\, \uparrow \rangle   \right), 
\nonumber\\
&& \nonumber \\
| t_{3,z} \rangle   &=& \frac{1}{\sqrt{2}}\left(
      |\uparrow\, \downarrow\, \uparrow\, \downarrow \rangle 
    - |\downarrow\, \uparrow\, \downarrow\, \uparrow \rangle     \right),
\nonumber \\
&& \nonumber \\
| d_0 \rangle   &=& \frac{1}{\sqrt{6}}\left(
      |\uparrow\, \downarrow\, \uparrow\, \downarrow \rangle 
    + |\downarrow\, \uparrow\, \downarrow\, \uparrow \rangle
    + |\uparrow\, \uparrow\, \downarrow\, \downarrow \rangle 
    + |\downarrow\, \downarrow\, \uparrow\, \uparrow \rangle \right. 
\nonumber \\
&& \nonumber \\
&& \left. \;\;\;\;\;
    + |\downarrow\, \uparrow\, \uparrow\, \downarrow \rangle 
    + |\uparrow\, \downarrow\, \downarrow\, \uparrow \rangle   \right), 
\nonumber \\
&& \nonumber \\
| d_\alpha \rangle   &=& \frac{\lambda_\alpha}{2\sqrt{2}}\left(
  \pm |\downarrow\, \uparrow\, \uparrow\, \uparrow \rangle 
  \pm |\uparrow\, \downarrow\, \uparrow\, \uparrow \rangle
  \pm |\uparrow\, \uparrow\, \downarrow\, \uparrow \rangle 
  \pm |\uparrow\, \uparrow\, \uparrow\, \downarrow \rangle \right. 
\nonumber \\
&& \left. \right. \nonumber \\
&& \left. \;\;\;\;\;\;
    + |\uparrow\, \downarrow\, \downarrow\, \downarrow \rangle 
    + |\downarrow\, \uparrow\, \downarrow\, \downarrow \rangle 
    + |\downarrow\, \downarrow\, \uparrow\, \downarrow \rangle 
    + |\downarrow\, \downarrow\, \downarrow\, \uparrow \rangle   \right), 
\nonumber\\
&& \nonumber \\
| d_z \rangle   &=& \frac{1}{\sqrt{2}}\left(
      |\downarrow\, \downarrow\, \downarrow\, \downarrow \rangle 
    - |\uparrow\, \uparrow\, \uparrow\, \uparrow \rangle     \right),
\nonumber \\
&& \nonumber \\
| d_2 \rangle   &=& \frac{1}{\sqrt{2}}\left(
      |\downarrow\, \downarrow\, \downarrow\, \downarrow \rangle 
    + |\uparrow\, \uparrow\, \uparrow\, \uparrow \rangle     \right),
\nonumber \\
\end{eqnarray}
where the upper and lower signs respectively refer to $\alpha =x$ and $y$,
$\lambda_x = 1$, and $\lambda_y = i$
In particular, the singlet states can also be written as\cite{singh99}
\begin{eqnarray}
 |s_0\rangle &=& -\frac{1}{\sqrt{3}}\left( [1,2][4,3] + [1,4][2,3] \right),
\nonumber \\
 && \label{s0-s1} \\
 |s_1\rangle &=& [1,2][4,3] - [1,4][2,3], 
\nonumber
\end{eqnarray}
where $[i,j]$ with $i,j = 1,2,3,4$ denotes that the spins $\bS^i$ and $\bS^j$
form a singlet, see Figs.~\ref{fig:model} and \ref{fig:singlets}.  
Note that $|s_0\rangle$ is even while $|s_1\rangle$ is odd under a $\pi/2$
rotation.

\section{Details: effective boson model}
\label{ap:coef}

Here, we quote the explicitly expressions of the coefficients $A^{ab}_\bk$, $B^{ab}_\bk$, 
$\xi^{abc}_{\bk}$, $\chi^{abcd}_\bk$, $\bar{\xi}^{ba}_\bp$, and $\bar{\chi}^{ab}_\bk$ 
[see Eqs.~\eqref{h20}--\eqref{h22}]:
\begin{eqnarray}
A^{ab}_\bk  &=& \left( E_{t1} - \mu \right)\left( \delta_{a,1}\delta_{b,1}
                             + \delta_{a,2}\delta_{b,2} \right)
\nonumber \\
&& \nonumber \\
  &+&   \left( E_{t3} - \mu \right)\delta_{a,3}\delta_{b,3}  + B^{ab}_\bk, 
\nonumber \\
&& \nonumber \\
B^{ab}_\bk    &=& N_0\sum_n \left[  g^{ab}_2(n)e^{i\bk\cdot\bn}  
                                + g^{ba}_2(n)e^{-i\bk\cdot\bn} \right], 
\nonumber \\
&& \nonumber \\
\xi^{abc}_\bk  &=& -iN^{1/2}_0\sum_n \left[  g^{abc}_3(n)e^{-i\bk\cdot\bn}  
                                  + \bar{g}^{bca}_3(n)e^{i\bk\cdot\bn} \right], 
\nonumber \\
&& \nonumber \\
\chi^{abcd}_\bk &=&  -\frac{1}{2}\sum_n \left[ g^{abcd}_4(n)e^{i\bk\cdot\bn}  
                                          + g^{cdab}_4(n)e^{-i\bk\cdot\bn} \right],
\nonumber \\
&& \nonumber \\
\bar{\xi}^{ab}_\bk  &=& N^{1/2}_0\sum_n \left[  g^{ab}_{so}(n)e^{-i\bk\cdot\bn}  
                                          + g^{ba}_{os}(n)e^{i\bk\cdot\bn} \right], 
\nonumber \\
&& \nonumber \\
\bar{\chi}^{ab}_\bk &=&  \sum_n \left[ g^{ab}_{ss}(n)e^{-i\bk\cdot\bn}  
                                   + g^{ba}_{ss}(n)e^{i\bk\cdot\bn} \right],
\label{coef-hboso01}
\end{eqnarray}
with $a,b,c,d = 1,2$, and $3$. $E_{t1}$ and $E_{t3}$ are the triplet eigenvalues
\eqref{eigenvalues} of the single plaquette Hamiltonian \eqref{ham-j1j2-plaq} 
and the $g$ coefficients are given by
\begin{eqnarray}
g^{ab}_{so}(n) &=& J_1\left( \bar{C}^2_aC^1_b + \bar{C}^3_aC^4_b \right)\delta_{n,1} 
\nonumber \\
&& \nonumber \\
             &+& J_1\left( \bar{C}^4_aC^1_b + \bar{C}^3_aC^2_b \right)\delta_{n,2} 
\nonumber \\
&& \nonumber \\
             &+& J_2\bar{C}^2_aC^4_b\left(\delta_{n,1}  + \delta_{n,-2} + \delta_{n,1-2}  \right) 
\nonumber \\
&& \nonumber \\
           &+& J_2\bar{C}^3_aC^1_b\left(\delta_{n,1}  + \delta_{n,2} + \delta_{n,1+2}  \right), 
\nonumber
\end{eqnarray}
\begin{eqnarray}
g^{abc}_3(n) &=& J_1\left( C^2_aD^1_{bc} + C^3_aD^4_{bc} \right)\delta_{n,1} 
\nonumber \\
&& \nonumber \\
           &+& J_1\left( C^4_aD^1_{bc} + C^3_aD^2_{bc} \right)\delta_{n,2} 
\nonumber \\
&& \nonumber \\
           &+& J_2C^2_aD^4_{bc}\left(\delta_{n,1}  + \delta_{n,-2} + \delta_{n,1-2}  \right) 
\nonumber \\
&& \nonumber \\
           &+& J_2C^3_aD^1_{bc}\left(\delta_{n,1}  + \delta_{n,2} + \delta_{n,1+2}  \right), 
\nonumber 
\end{eqnarray}
\begin{eqnarray}
\bar{g}^{abc}_3(n) &=& J_1\left( D^2_{ab}C^1_c + D^3_{ab}C^4_c \right)\delta_{n,1} 
\nonumber \\
&& \nonumber \\
            &+& J_1\left( D^4_{ab}C^1_c + D^3_{ab}C^2_c \right)\delta_{n,2} 
\nonumber \\
&& \nonumber \\
            &+& J_2D^2_{ab}C^4_c\left(\delta_{n,1}  + \delta_{n,-2} + \delta_{n,1-2}  \right) 
\nonumber \\
&& \nonumber \\
           &+& J_2D^3_{ab}C^1_c\left(\delta_{n,1}  + \delta_{n,2} + \delta_{n,1+2}  \right), 
\nonumber 
\end{eqnarray}
\begin{eqnarray}
g^{abcd}_4(n) &=& J_1\left( D^2_{ab}D^1_{cd} + D^3_{ab}D^4_{cd} \right)\delta_{n,1} 
\nonumber \\
&& \nonumber \\
            &+& J_1\left( D^4_{ab}D^1_{cd} + D^3_{ab}D^2_{cd} \right)\delta_{n,2} 
\nonumber \\
&& \nonumber \\
            &+& J_2D^2_{ab}D^4_{cd}\left(\delta_{n,1}  + \delta_{n,-2} + \delta_{n,1-2}  \right) 
\nonumber \\
&& \nonumber \\
           &+& J_2D^3_{ab}D^1_{cd}\left(\delta_{n,1}  + \delta_{n,2} + \delta_{n,1+2}  \right), 
\label{coef-hboso02} 
\end{eqnarray}
$g^{ab}_2(n) = g^{ab}_{so}(n)$ with the replacements $\bar{C} \leftrightarrow C$,
$g^{ab}_{os}(n) = g^{ab}_{so}(n)$ with the replacements $C \leftrightarrow \bar{C}$,
and $g^{ab}_{ss}(n) = g^{ab}_{so}(n)$ with $C \rightarrow \bar{C}$.
Here $n = 1,2$ corresponds to the nearest-neighbor vectors
\eqref{tauvectors} and the $C$, $\bar{C}$, and $D$ coefficients are 
shown in Eq.~\eqref{spin-bondop3-coef}.

\section{Diagonalization harmonic Hamiltonian}
\label{ap:diag}

In this section, we briefly summarize the analytical procedure
used to diagonalize the triplet sector of the harmonic Hamiltonian
\eqref{ham-harmonic-matrix}. In order to deal with such a 
$6\times 6$ problem, we follow the procedure described in 
Refs.~\onlinecite{colpa78} and \onlinecite{blaizot}. 
It should be mentioned that we have recently employed this scheme 
to diagonalize a similar $4\times 4$ problem.\cite{doretto12-2}  

Since we are considering a bosonic system, instead of $\hat{H}_\bk$
[see Eq.~\eqref{ham-harmonic-matrix}], we should diagonalize
\begin{equation}
 \hat{I}_B\hat{H}_\bk ,
 \;\;\;\;\;\;\; {\rm with}
 \;\;\;\;\;\;\;
 \hat{I}_B = \left( \begin{array}{cc}
                      \hat{I} & 0 \\ 0 & -\hat{I}
                      \end{array}  \right),
\label{aux-ham}
\end{equation}
where $\hat{I}$ is the $3\times 3$ identity matrix. It is easy to show
that the (positive) eigenvalues of the matrix \eqref{aux-ham} are
(roots of a cubic polynomial)
\begin{eqnarray}
 \Omega_{1/2,\bk} &=& \left[  -\frac{1}{3}a_{2,\bk} - {\rm Re}(S_\bk) \mp
                                  \sqrt{3}{\rm Im}(S_\bk) \right]^{1/2},
\nonumber \\
&& \\
 \Omega_{3,\bk} &=& \left[  -\frac{1}{3}a_{2,\bk} + 2{\rm Re}(S_\bk)  \right]^{1/2},
\nonumber
\end{eqnarray}
where
\begin{eqnarray}
 S_\bk &=& \left( R_\bk + i\sqrt{D_\bk} \right)^{1/3},
\;\;\;\;\;\;\;\;\;\;\;
 D_\bk = -Q^3_\bk - R^2_\bk,
\nonumber  \\
 Q_\bk &=& \frac{1}{9} \left( 3a_{1,\bk} - a^2_{2,\bk} \right), 
\\
 R_\bk &=& \frac{1}{54}\left( 9a_{2,\bk}a_{1,\bk} - 27a_{0,\bk} - 2a^3_{2,\bk} \right).
\nonumber
\end{eqnarray}
The coefficients $a_{i,\bk}$ read
\begin{eqnarray}
a_{0,\bk} &=& \left( A^{11}_\bk - B^{11}_\bk \right) \left( A^{22}_\bk - B^{22}_\bk \right)
            \left( A^{33}_\bk - B^{33}_\bk \right)\left[ \right.
\nonumber \\
&& \nonumber \\
        && \left. 4(B^{12}_\bk)^2\left( A^{33}_\bk + B^{33}_\bk \right) 
                + 4(B^{23}_\bk)^2\left( A^{11}_\bk + B^{11}_\bk \right)  \right.
\nonumber \\
&& \nonumber \\
        &+& \left. 4(B^{13}_\bk)^2\left( A^{22}_\bk + B^{22}_\bk \right)
                -  16B^{12}_\bk B^{13}_\bk B^{23}_\bk  \right.
\nonumber \\
&& \nonumber \\
        &-& \left.
             \left( A^{11}_\bk + B^{11}_\bk \right) \left( A^{22}_\bk + B^{22}_\bk \right)
             \left( A^{33}_\bk + B^{33}_\bk \right)  \right],
\nonumber \\
&& \nonumber \\ 
a_{1,\bk} &=&     \omega^2_{1,\bk}\omega^2_{2,\bk} + \omega^2_{1,\bk}\omega^2_{3,\bk}
              + \omega^2_{2,\bk}\omega^2_{3,\bk}
\nonumber \\
&& \nonumber \\
        &-& 4(B^{12}_\bk)^2\left( A^{11}_\bk - B^{11}_\bk \right) \left( A^{22}_\bk - B^{22}_\bk \right)
\nonumber \\
&& \nonumber \\
        &-& 4(B^{23}_\bk)^2\left( A^{22}_\bk - B^{22}_\bk \right) \left( A^{33}_\bk - B^{33}_\bk \right)
\nonumber \\
&& \nonumber \\
        &-& 4(B^{13}_\bk)^2\left( A^{11}_\bk - B^{11}_\bk \right) \left(A^{33}_\bk - B^{33}_\bk \right),
\nonumber \\
&& \nonumber \\
a_{2,\bk} &=&  - \left(\omega^2_{1,\bk} + \omega^2_{2,\bk} + \omega^2_{3,\bk} \right),
\end{eqnarray}
where
$
  \omega^2_{i,\bk} = (A^{ii}_\bk)^2 - (B^{ii}_\bk)^2,
$
with $i=1,2$, and $3$, and $A^{ab}_\bk$ and $B^{ab}_\bk$ [see Eq.~\eqref{coef-hboso01}]
are respectively the elements of the $3\times 3$ Hermitian 
$\hat{A}_\bk$ and $\hat{B}_\bk$ [see Eq.~\eqref{matrix-AB}].

The determination of the Bogoliubov coefficients $u^{ab}_\bk$ and
$v^{ab}_\bk$, the elements of the $3\times 3$ matrices $\hat{U}_\bk$
and $\hat{V}_\bk$, is quite involved. Using the properties of the
matrix $\hat{M}_\bk$ (see Sec.~5 from Ref.~\onlinecite{colpa78}) and
after some lengthy algebra, it is possible to show that   
\begin{eqnarray}
u^{jb}_\bk &=& \frac{\mu_{jb,\bk}}{G_{b,\bk}}
             \left(B^{jj}_\bk - A^{jj}_\bk  - \Omega_{b,\bk} \right)
             \left(\Omega_{b,\bk}  +  A^{33}_\bk  - B^{33}_\bk  \right),
\nonumber \\
&& \nonumber \\
u^{3b}_\bk &=& -\frac{i}{G_{b,\bk}} \left[  \left( \Omega_{b,\bk} + A^{33}_\bk \right)\nu_{b,\bk}
                   + 2\lambda_{b,\bk}B^{23}_\bk\left( A^{22}_\bk  - B^{22}_\bk \right) 
                   \right.
\nonumber \\
&& \nonumber \\
                   &+& \left.  2\mu_{b,\bk}B^{13}_\bk\left( A^{11}_\bk  - B^{11}_\bk \right)
                     \right],
\nonumber \\
&& \nonumber \\
v^{jb}_\bk &=& \frac{\mu_{jb,\bk}}{G_{b,\bk}}
             \left(\Omega_{b,\bk}  - A^{jj}_\bk  + B^{jj}_\bk \right)
             \left(\Omega_{b,\bk}  + A^{33}_\bk  - B^{33}_\bk \right),
\nonumber \\
&& \nonumber \\
v^{3b}_\bk &=& \frac{i}{G_{b,\bk}} \left[  B^{33}_\bk \nu_{b,\bk}
                   + 2\lambda_{b,\bk}B^{23}_\bk\left( A^{22}_\bk  - B^{22}_\bk \right) 
                   \right.
\nonumber \\
&& \nonumber \\
         &+& \left.  2\mu_{b,\bk}B^{13}_\bk\left( A^{11}_\bk  - B^{11}_\bk \right)
                     \right],
\end{eqnarray}
where $j=1,2$, 
\begin{eqnarray}
 \mu_{1b,\bk} &=& 2B^{21}_\bk B^{23}_\bk \left( A^{22}_\bk  - B^{22}_\bk \right) 
                       + B^{13}_\bk \left( \Omega^2_{b,\bk} - \omega^2_{2,_\bk} \right), 
 \nonumber \\
&& \nonumber \\
 \mu_{2b,\bk} &=& 2B^{21}_\bk B^{13}_\bk \left( A^{11}_\bk  - B^{11}_\bk \right) 
                            + B^{23}_\bk \left( \Omega^2_{b,\bk} - \omega^2_{1,\bk} \right), 
\nonumber \\
&& \nonumber \\
 \nu_{b,\bk} &=& \left( \Omega^2_{b,\bk} - \omega^2_{1,\bk} \right)
               \left( \Omega^2_{b,\bk} - \omega^2_{2,\bk} \right)
\nonumber \\
&& \nonumber \\
           &-& 4(B^{21}_\bk)^2 \left( A^{11}_\bk  - B^{11}_\bk \right) 
                             \left( A^{22}_\bk  - B^{22}_\bk \right),
\nonumber 
\end{eqnarray}
and
\begin{eqnarray}
 G^2_{b,\bk} &=& 4\Omega_{b,\bk}\left( \Omega_{b,\bk} + A^{33}_\bk - B^{33}_\bk \right)^2 
                 \left[  \mu^2_{b,\bk}\left( A^{11}_\bk - B^{11}_\bk \right) \right.   
\nonumber \\
&& \left. \right. \nonumber \\
           &+&  \left. \lambda^2_{b,\bk}\left( A^{22}_\bk - B^{22}_\bk \right) \right]  
\nonumber \\
&& \nonumber \\
           &+& \nu_{b,\bk}\left( \Omega_{b,\bk} + A^{33}_\bk - B^{33}_\bk \right)\left[ 
               \nu_{b,\bk}\left( \Omega_{b,\bk} + A^{33}_\bk + B^{33}_\bk \right)
               \right.
\nonumber \\ 
&& \left. \right. \nonumber \\
           &+& \left.
               4\lambda_{b,\bk}B^{23}_\bk\left( A^{22}_\bk  - B^{22}_\bk \right) 
            +  4\mu_{b,\bk}B^{13}_\bk \left( A^{11}_\bk  - B^{11}_\bk \right) \right],
\nonumber 
\end{eqnarray}
with $b=1,2$, and $3$.

\section{Dimerized columnar and staggered VBSs -- harmonic approximation}
\label{ap:col-stag}

In this section, we study the dimerized columnar [Fig.~\ref{fig:dimer}(a)] and 
staggered [Fig.~\ref{fig:dimer}(b)] VBS phases of the $J_1$--$J_2$ model within 
the (dimer) bond--operator formalism\cite{sachdev90} at the harmonic
approximation. We only quote the main results and refer the
reader to Secs.~II and III from Ref.~\onlinecite{doretto12} for more  
details. 

\begin{figure}[t]
\centerline{\includegraphics[width=7.1cm]{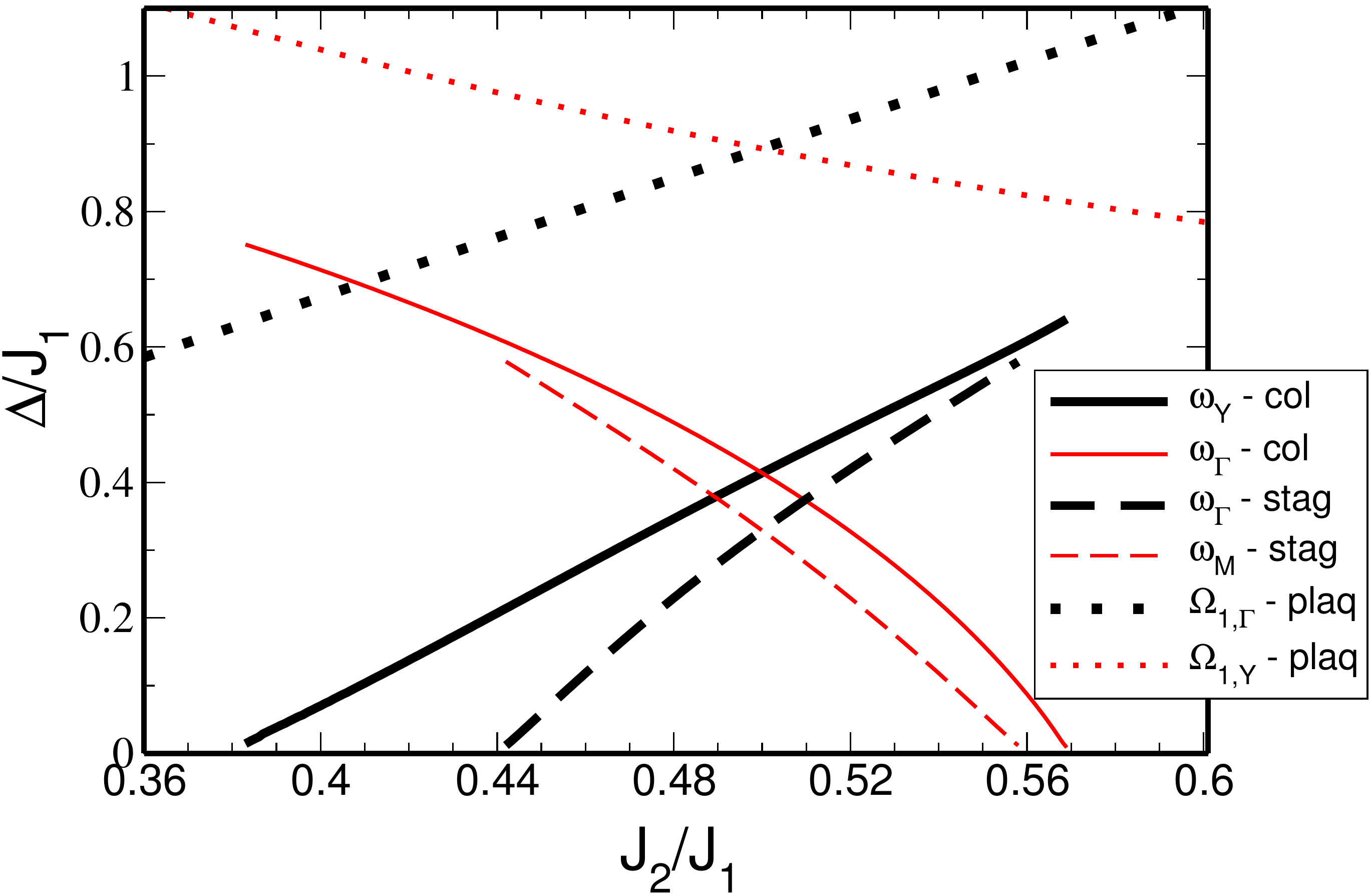}}
\caption{(Color online) Excitation gaps of the dimerized columnar (solid lines)
         and staggered (dashed lines) valence bond solid phases at
         the harmonic approximation.
         $\omega_{\rm Y}$ (thick solid black line) and $\omega_\Gamma$ (thin solid red line) 
         are the energy of the triplet excitations [Eq.~\eqref{disp-dimer}] of 
         the columnar VBS at the ${\rm Y} = (0,\pi)$ and $\Gamma = (0,0)$
         points of the dimerized Brillouin zone. 
         Similar for the staggered VBS, $\omega_\Gamma$ (thick dashed black line) and 
         $\omega_{\rm M}$ (thin dashed red line) are the energy of the triplet
         excitations at the $\Gamma = (0,0)$ and ${\rm M} = (\pi,\pi)$
         points of the dimerized Brillouin zone. 
         The dotted lines are the gap of the plaquette valence bond
         solid (triplet excitations, harmonic approximation) as shown 
         in Fig.~\ref{fig:gap}(a).
}
\label{fig:gap-col-stag}
\end{figure}

The effective model [the equivalent of Eq.~\eqref{ham-bond}] in terms 
of the boson triplet operators $t_{\bk\alpha}$ with $\alpha = x,y,z$ reads
\begin{eqnarray}
\mathcal{H} &=& -3J_1N/8 - \mu N(N_0 - 1)/2
\nonumber \\
&& \nonumber \\
           &+& \sum_\bk \left[ A_\bk t^\dagger_{\bk\alpha}t_{\bk\alpha}
                  + \frac{1}{2}B_\bk \left( t^\dagger_{\bk\alpha}t^\dagger_{\bk\alpha}
                  + {\rm H.c.}\right) \right]
\nonumber\\
&& \nonumber \\
          &+& \frac{1}{2\sqrt{N'}}\epsilon_{\alpha\beta\lambda}\sum_{\bp,\bk}\xi_{\bk-\bp}
                  \; t^\dagger_{\bk-\bp\alpha}t^\dagger_{\bp\beta}t_{\bk\lambda} + {\rm H.c.}
\nonumber \\
&& \nonumber \\
          &+& \frac{1}{2N'}\epsilon_{\alpha\beta\lambda}\epsilon_{\alpha\mu\nu}
                  \sum_{\bq,\bp,\bk} \gamma_\bk \;
                   t^\dagger_{\bp+\bk\beta}t^\dagger_{\bq-\bk\mu}t_{\bq\nu}t_{\bp\lambda}.
\label{boso-dimer}
\end{eqnarray}
Here, $N' = N/2$ with $N$ being the number of sites of the original
square lattice, the momentum sum runs over the dimerized Brillouin zone, 
\begin{eqnarray}
  A_\bk &=& \frac{1}{4}J_1 - \mu + B_\bk, 
\nonumber \\
&& \nonumber \\
 B_\bk &=& \frac{1}{2}N_0\left[ -J_1\cos(2k_x) +2 (J_1 - J_2)\cos(k_y)  \right.
\nonumber \\
&& \left. \right. \nonumber \\
         &-& \left.  J_2\cos(2k_x + k_y) - J_2\cos(2k_x - k_y) \right],       
\nonumber \\ 
&& \nonumber \\
 \xi_\bk    &=& -\sqrt{N_0}\left[J_1\sin(2k_x) + J_2\sin(2k_x + k_y)  \right.
\nonumber \\
&& \left. \right. \nonumber \\
         &+& \left.            J_2\sin(2k_x - k_y) \right],
\nonumber \\
&& \nonumber\\
 \gamma_\bk &=& -\frac{1}{2}\left[  J_1\cos(2k_x) + 2(J_1 + J_2)\cos k_y \right.
\nonumber \\
&& \left. \right. \nonumber \\
                    &+& \left. J_2\cos(2k_x + k_y)  +  J_2\cos(2k_x - k_y) \right]
\end{eqnarray}
for the columnar VBS, and 
\begin{eqnarray}
 B_\bk &=& \frac{1}{2}N_0\left[ (2J_2 -J_1)(\cos k_x + \cos k_y )
               - J_1\cos(k_x - k_y)\right],    
\nonumber  \\ 
&& \nonumber \\
 \xi_\bk    &=& \sqrt{N_0}J_1\left[\sin k_y + \sin(k_y - k_x)
                - \sin(k_x) \right],
\\
&& \nonumber  \\
 \gamma_\bk &=& -\frac{1}{2}\left[ (2J_2 + J_1)(\cos k_x + \cos k_y) 
                             + J_1\cos(k_x - k_y) \right]
\nonumber 
\end{eqnarray}
for the staggered VBS. 
In deriving Eq.~\eqref{boso-dimer}, we considered the following nearest--neighbor
vectors: 
$\taub_1 = 2a\hat{x} = {\bf a}_1$ and $\taub_2 = a\hat{y} = {\rm a}_2$ (columnar) 
and 
$\taub_1 = a\hat{x} = {\bf a}_1$ and $\taub_2 = a\hat{y} = {\rm a}_2$ (staggered), 
see Fig.~\ref{fig:dimer}, and we set $a=1$. 
Similar to the plaquette phase, the parameter
$N^{1/2}_0$ is the average value of the singlet operator $s_i$
while $\mu$ is the Lagrange multiplier that enforce (on average) 
the constraint on the total number of bosons per site (dimerized lattice).

Within the harmonic approximation, the Hamiltonian \eqref{boso-dimer}
can be diagonalized, and therefore 
one finds that the ground state energy is given by
\begin{equation}
  E_{EGS} = -\frac{3}{8}J_1NN_0 - \frac{1}{2}\mu N(N_0 - 1)
     + \frac{3}{2}\sum_\bk \left(\omega_\bk - A_\bk \right)
\label{egs-harmonic}
\end{equation}
while the energy of the triplet excitations assume the form
\begin{equation}
 \omega_\bk = \sqrt{A^2_\bk - B^2_\bk}.
\label{disp-dimer}
\end{equation}
After self--consistently calculating $N_0$ and $\mu$, we find the
behaviour of the ground state energy [Fig.~\ref{fig:egs}(b)] and the
excitation gaps (Fig.~\ref{fig:gap-col-stag}) in terms of $J_2/J_1$.  
Recall that for the columnar VBS phase, 
the ${\rm Y} = (0,\pi)$ and $\Gamma = (0,0)$ vectors
correspond to the $(\pi,\pi)$ and $(0,0)$ vectors of the 
original (nondimerized) square lattice.\cite{kotov99}

\section{Details: cubic--quartic approximation}
\label{ap:cubic}

The renormalized cubic vertices $\Gamma^{abc}_{1/2,\bk,\bp}$ and 
$\Gamma^{ab}_{1s/3s/4s,\bk,\bp}$ [see Fig.~\ref{fig:vertices}(a)
and Eqs.~\eqref{h30-v2} and \eqref{h21-v2}]  
in terms of the Bogoliubov coefficients $u^{ab}_\bk$ and $v^{ab}_\bk$
are given by 
\begin{eqnarray}
&& \Gamma^{abc}_{1,\bk,\bp} = \sum_{\bar{a}\bar{b}\bar{c}}
 \left(  \xi^{\bar{a}\bar{b}\bar{c}}_{\bp-\bk} - \xi^{\bar{b}\bar{a}\bar{c}}_{-\bp}  \right)
  u^{\bar{a}a\,\dagger}_{\bk-\bp}u^{\bar{b}b\,\dagger}_\bp u^{\bar{c}c}_\bk 
\nonumber \\
&& \nonumber \\
 &+&  \left(  \xi^{\bar{a}\bar{b}\bar{c}}_{\bk-\bp} - \xi^{\bar{b}\bar{a}\bar{c}}_{\bp}\right) 
   v^{\bar{a}a}_{\bp-\bk}v^{\bar{b}b}_{-\bp}v^{\bar{c}c\,\dagger}_{-\bk}
   + \left(  \xi^{\bar{b}\bar{c}\bar{a}}_{\bp} - \xi^{\bar{c}\bar{b}\bar{a}}_{-\bk}  \right)
   u^{\bar{a}a\,\dagger}_{\bk-\bp}v^{\bar{b}b}_{-\bp}u^{\bar{c}c}_\bk
\nonumber \\
&& \nonumber \\
 &+& \left(  \xi^{\bar{c}\bar{a}\bar{b}}_\bk - \xi^{\bar{a}\bar{c}\bar{b}}_{\bp-\bk}  \right)
   u^{\bar{a}a\,\dagger}_{\bk-\bp}v^{\bar{b}b}_{-\bp}v^{\bar{c}c\,\dagger}_{-\bk}
 + \left(  \xi^{\bar{c}\bar{a}\bar{b}}_{-\bk} - \xi^{\bar{a}\bar{c}\bar{b}}_{\bk-\bp}\right)
  v^{\bar{a}a}_{\bp-\bk}u^{\bar{b}b\,\dagger}_\bp u^{\bar{c}c}_\bk
\nonumber \\
&& \nonumber \\
 &+& \left(  \xi^{\bar{b}\bar{c}\bar{a}}_{-\bp} - \xi^{\bar{c}\bar{b}\bar{a}}_{\bk}  \right)
   v^{\bar{a}a}_{\bp-\bk}u^{\bar{b}b\,\dagger}_\bp v^{\bar{c}c\,\dagger}_{-\bk},
\label{gammas}
\end{eqnarray}
$\Gamma^{abc}_{2,\bk,\bp} = \Gamma^{abc}_{1,\bk,\bp}$ with the
replacement $u^{\bar{c}c}_\bk \leftrightarrow v^{\bar{c}c}_\bk$,  
\begin{eqnarray}
\Gamma^{ab}_{1s,\bk,\bp} &=& \sum_{cd}
     \left(  \bar{\xi}^{cd}_\bp + \bar{\xi}^{dc}_{\bk-\bp}\right)  
   u^{ca\,\dagger}_{\bk-\bp}v^{db\,\dagger}_{\bp} 
\nonumber \\
&& \nonumber \\
 &+& \bar{\xi}^{cd\, *}_{-\bp}u^{ca\,\dagger}_{\bk-\bp}u_{db,-\bp}
   + \bar{\xi}^{dc\, *}_{\bp-\bk}v_{ca,\bp-\bk}v^{db\,\dagger}_\bp,
\\
&& \nonumber \\
\Gamma^{ab}_{3s,\bk,\bp} &=& \sum_{cd}
         \bar{\xi}^{cd}_\bp u^{ca\,\dagger}_{\bk-\bp}u^{db\,\dagger}_\bp
      +  \bar{\xi}^{cd\, *}_{-\bp}u^{ca\,\dagger}_{\bk-\bp}v^{db}_{-\bp},
\nonumber 
\end{eqnarray}
and $\Gamma^{ab}_{4s,\bk,\bp} = \Gamma^{ab}_{3s,\bk,\bp}$ with the
replacement $u \leftrightarrow v$.

The components of the normal triplet $\Sigma_a(\bk,\omega)$ 
[Eq.~\eqref{self-triplet}] and singlet $\Sigma_s(\bk,\omega)$ 
[Eq.~\eqref{self-singlet}] self--energies read
\begin{eqnarray}
   \Sigma^{(b1)}_a(\bk,\omega) &=& \frac{1}{N'}\sum_{b,c}\sum_\bp
      \frac{ |\Gamma^{cba}_{1,\bk,\bp}|^2}{\omega - \Omega_{b,\bp}
                       - \Omega_{c,\bk-\bq} + i\delta},
\nonumber \\
&& \nonumber \\
    \Sigma^{(b4)}_a(\bk,\omega) &=& -\frac{1}{N'}\sum_{b,c}\sum_\bp
       \frac{|\Gamma^{cba}_{2,-\bk,-\bp}|^2}{\omega + \Omega_{b,-\bp}
                       + \Omega_{c,\bp-\bk} - i\delta},
\nonumber \\ 
&& \label{triplet-self-energy}\\
    \Sigma^{(c1)}_a(\bk,\omega) &=& \frac{1}{N'}\sum_b\sum_\bp
      \frac{ |\Gamma^{ab}_{1s,\bp,\bp-\bk}|^2}
              {\omega - \Omega_s - \Omega_{b,\bk-\bq} + i\delta},
\nonumber \\
&& \nonumber \\
    \Sigma^{(c3)}_a(\bk,\omega) &=& -\frac{1}{N'}\sum_b\sum_\bp
       \frac{ |\Gamma^{ab}_{4s,-\bp,\bk-\bp}|^2}
               {\omega + \Omega_s + \Omega_{b,\bp-\bk} - i\delta}.
\nonumber
\end{eqnarray}
with $a = 1,2,3$ and
\begin{eqnarray}
   \Sigma^{(d2)}_s(\bk,\omega) &=& \frac{3}{N'}\sum_{a,b}\sum_\bp
      \frac{ \Gamma^{ab}_{3s,\bk,\bp}\left(  \Gamma^{ab\,*}_{3s,\bk,\bp} 
                                       + \Gamma^{ba\*}_{3s,\bk,\bk-\bp}  \right)}
              {\omega - \Omega_{b,\bp} - \Omega_{a,\bk-\bq} + i\delta},
\nonumber \\
&& \label{singlet-self-energy}\\
    \Sigma^{(d3)}_s(\bk,\omega) &=& -\frac{3}{N'}\sum_{a,b}\sum_\bp
       \frac{\Gamma^{ab}_{4s,\bk,\bp}\left( \Gamma^{ab\,*}_{4s,\bk,\bp}  
                                      + \Gamma^{ba\,*}_{4s,\bk,\bp-\bk} \right)}
               {\omega + \Omega_{b,-\bp} + \Omega_{a,\bp-\bk} - i\delta}.
\nonumber
\end{eqnarray}

The coefficients of the quartic terms $\mathcal{H}_{40}$ and $\mathcal{H}_{22}$
within the Hartree--Fock approximation [Eqs.~\eqref{h40-v2} and \eqref{h22-v2}]
are given by
\begin{eqnarray}
  A^{HF}_{ab,\bp} &=& \frac{3}{N'}\sum_{c,\bar{a},\bar{b},\bar{c},\bar{d}}\sum_\bk
    \chi^{\bar{a}\bar{b}\bar{c}\bar{d}}_\bk\left(
    u^{\bar{a}a\,\dagger}_{\bp}v^{\bar{b}b\,\dagger}_{-\bp}u^{\bar{c}c}_{-\bp+\bk}v^{\bar{d}c}_{\bp-\bk}
   \right. 
\nonumber \\
&& \left. \right. \nonumber \\
  &+& \left. 
    v^{\bar{a}b\,\dagger}_{-\bp}u^{\bar{b}a\,\dagger}_{\bp}u^{\bar{c}c}_{\bp+\bk}v^{\bar{d}c}_{-\bp-\bk}
  - v^{\bar{a}b\,\dagger}_{-\bp}v^{\bar{b}c\,\dagger}_{-\bp-\bk}v^{\bar{c}a}_{-\bp}v^{\bar{d}c}_{-\bp-\bk}
     \right.
\nonumber \\
&& \left. \right. \nonumber \\
  &-& \left. 
    u^{\bar{a}a\,\dagger}_{\bp}v^{\bar{b}c\,\dagger}_{\bp-\bk}u^{\bar{c}b}_{\bp}v^{\bar{d}c}_{\bp-\bk}
  + v^{\bar{a}c\,\dagger}_{-\bp+\bk}u^{\bar{b}c\,\dagger}_{\bp-\bk}u^{\bar{c}b}_{\bp}v^{\bar{d}a}_{-\bp}
     \right.
\nonumber \\
&& \left. \right. \nonumber \\
  &+& \left. 
    v^{\bar{a}c\,\dagger}_{\bp+\bk}u^{\bar{b}c\,\dagger}_{-\bp-\bk}v^{\bar{c}a}_{-\bp}u^{\bar{d}b}_{\bp}
  - v^{\bar{a}c\,\dagger}_{-\bp+\bk}v^{\bar{b}b\,\dagger}_{-\bp}v^{\bar{c}c}_{-\bp+\bk}v^{\bar{d}a}_{-\bp}
     \right.
\nonumber \\
&& \left. \right. \nonumber \\
  &-& \left. 
    v^{\bar{a}c\,\dagger}_{\bp+\bk}u^{\bar{b}a\,\dagger}_{\bp}v^{\bar{c}c}_{\bp+\bk}u^{\bar{d}b}_{\bp}
   \right),
\end{eqnarray}
\begin{eqnarray}
  B^{HF}_{ab,\bp} &=& \frac{3}{N'}\sum_{c,\bar{a},\bar{b},\bar{c},\bar{d}}\sum_\bk
    \chi^{\bar{a}\bar{b}\bar{c}\bar{d}}_\bk\left(
    u^{\bar{a}a\,\dagger}_{\bp}u^{\bar{b}b\,\dagger}_{-\bp}u^{\bar{c}c}_{-\bp+\bk}v^{\bar{d}c}_{\bp-\bk}
   \right. 
\nonumber \\
&& \left. \right. \nonumber \\
  &-& \left. 
    u^{\bar{a}a\,\dagger}_{\bp}v^{\bar{b}c\,\dagger}_{\bp-\bk}v^{\bar{c}b}_{\bp}v^{\bar{d}c}_{\bp-\bk}
  + v^{\bar{a}c\,\dagger}_{\bp+\bk}u^{\bar{b}c\,\dagger}_{-\bp-\bk}v^{\bar{c}a}_{-\bp}v^{\bar{d}b}_{\bp}
     \right.
\nonumber \\
&& \left. \right. \nonumber \\
  &-& \left. 
    v^{\bar{a}c\,\dagger}_{-\bp+\bk}u^{\bar{b}b\,\dagger}_{-\bp}v^{\bar{c}c}_{-\bp+\bk}v^{\bar{d}a}_{-\bp}
   \right),
\end{eqnarray}
\begin{eqnarray}
  A^{HF}_{s,\bp} &=& \frac{3}{N'}\sum_{a}\sum_\bk
    \bar{\chi}^{\bar{a}\bar{b}}_\bk
    v^{\bar{a}a\,\dagger}_{\bp-\bk}v^{\bar{b}a}_{\bp-\bk},
\nonumber \\
&&  \\
  B^{HF}_{s,\bp} &=& \frac{3}{N'}\sum_{a}\sum_\bk
    \bar{\chi}^{\bar{a}\bar{b}}_\bk
    u^{\bar{a}a}_{\bp+\bk}v^{\bar{b}a}_{-\bp-\bk}.
\nonumber 
\end{eqnarray}

Finally, it is possible to show\cite{zhito09} that only the 
cubic vertices $(2)$ and $(4s)$ in Fig.~\ref{fig:vertices}(a) 
contribute to the ground state energy and therefore, we have 
\begin{eqnarray}
  E^{(3)}_{EGS} &=& -\frac{1}{N'}\sum_{a,b,c}\sum_{\bk,\bp}
       \frac{|\Gamma^{abc}_{2,-\bk,-\bp}|^2}
             {\Omega_{c,\bk} + \Omega_{b,\bp} + \Omega_{a,\bp-\bk}}
\nonumber \\
&& \nonumber \\
 &-&    \frac{3}{N'}\sum_{a,b,c}\sum_{\bk,\bp}
       \frac{\Gamma^{ab}_{4s,\bk,\bp}\left( \Gamma^{ab\,*}_{4s,\bk,\bp}  
                                      + \Gamma^{ba\,*}_{4s,\bk,\bp-\bk} \right)}
               {\Omega_s + \Omega_{b,-\bp} + \Omega_{a,\bp-\bk}}.
\;\;\;\;\;
\label{egs-cubic}
\end{eqnarray}
The correction to the ground state energy due to the quartic terms read
\begin{eqnarray}
 E^{(4)}_{EGS} &=& \frac{9}{N'}\sum_{a,b,\bar{a},\bar{b},\bar{c},\bar{d}}\sum_{\bp,\bk}
    \chi^{\bar{a}\bar{b}\bar{c}\bar{d}}_\bk\left(
    v^{\bar{a}a\,\dagger}_{\bp+\bk}v^{\bar{b}b\,\dagger}_\bp v^{\bar{c}a}_{\bp+\bk}v^{\bar{d}b}_\bp
   \right. 
\nonumber \\
&& \left. \right. \nonumber \\
  &+& \left. 
    v^{\bar{a}a\,\dagger}_{\bp+\bk}u^{\bar{b}a\,\dagger}_{-\bp-\bk} u^{\bar{c}b}_{-\bp}v^{\bar{d}b}_\bp
     \right).
\end{eqnarray}



\begin{thebibliography}{}
\bibitem{comment2}
The fact that $\bar{\Gamma}_{1,\Gamma}$, $\bar{\Gamma}_{1,X}$, and
$\bar{\Gamma}_{s,\Gamma}$ are not strictly zero and that $\bar{\Gamma}_{s,X}$
has small oscillations in Fig.~\ref{fig:decay2} might be related to the 
numerical procedure employed to evaluate the integrals in 
Eqs.~\eqref{triplet-self-energy} and \eqref{singlet-self-energy}.
\end{thebibliography}
\end{document}